\documentclass[11pt]{article}
\usepackage{jcappub}
\usepackage{graphicx}
\usepackage{caption}
\usepackage[utf8x]{inputenc}
\usepackage{color}
\usepackage{subfig}
\usepackage{multirow}
\usepackage{slashed}
\usepackage{epstopdf}
\usepackage{subfig}
\usepackage{bmpsize}

\title{Some Implications of the Leptonic Annihilation of Dark Matter: Possible Galactic Radio Emission Signatures and the Excess Radio Flux of Extragalactic Origin}

\author[a,b]{Elaine C. F. S. Fortes}%
\affiliation[a]{Universidade Federal do Pampa\\
	Rua Luiz Joaquim de Sá Brito, s/n, Promorar,\\ Itaqui - RS, 97650-000,
	Brazil}
\affiliation[b]{Instituto Nacional de Pesquisas Espaciais\\
	Av. dos Astronautas, 1758 - Jardim da Granja, \\São Jose dos Campos, SP, 01506-000, Brazil}

\author[b]{Oswaldo D. Miranda}%

\author[c]{Floyd W. Stecker}%
\affiliation[c]{Astrophysics Science Division, \\ NASA Goddard Space Flight Center, Greenbelt, MD 20771
}

\author[b]{Carlos A. Wuensche}%

\emailAdd{elainefortes@unipampa.edu.br}
\emailAdd{oswaldo.miranda@inpe.br}
\emailAdd{floyd.w.stecker@nasa.gov}
\emailAdd{ca.wuensche@inpe.br}

\abstract{We give theoretical predictions for the radio emission of a dark matter candidate annihilating into 2-lepton and 4-lepton final states. We then compare our results with the known radio measurements of the sky temperature as a function of the frequency. In particular, we calculate the radio emission for some dark matter candidates annihilating into intermediate bosons that subsequently decay into a 4-lepton channel with a thermal annihilation cross-section. We show that within the range of frequencies from $20\,{\rm MHz}$ to $5\,{\rm GHz}$, this channel can produce a stronger signature than direct annihilation into leptons.}

	
\begin{document}	
\maketitle

\section{Introduction}
\label{sec:intro}

A combination of cosmological and astrophysical observations dating back decades has provided strong evidence for the existence of dark matter (DM). This evidence includes data from rotational curves of galaxies, the dynamics of galaxy-galaxy interactions, spatial temperature fluctuations in the cosmic microwave background (CMB) radiation, and gravitational lensing. The luminous matter observed in spiral galaxies is not abundant by itself to account for the rotation curves of the galaxies (see, e.g. \cite{Rubin:1982}). The CMB fluctuation results infer that DM component  of the universe is $\Omega_{\rm dm} h^{2}=0.120\pm 0.001$ while the baryonic component is $\Omega_{\rm b} h^{2}=0.0224\pm 0.0001$ \cite{Planck2018}. Observations of gravitational lensing phenomena also imply the presence of the DM, showing that evidence of a significant amount of mass where nothing is observed optically (see, e.g., \cite{Clowe_2006,Massey:2007wb}).

We note that strong limits on the direct interaction of DM particles with standard model particles have been obtained in the laboratory~\cite{Aprile:2019}--\cite{Fu:2016ega}).
Other attempts to search for DM were performed by looking for the secondary $\gamma$-rays produced by DM annihilation and pion decay, as well as from processes such as internal bremsstrahlung~\cite{Bringmann:2012ez, Bringmann:2012vr} and synchrotron radiation \cite{Ishiwata:2008qy}. The secondary radiation consists of three kinds: (1) radio waves due to synchrotron radiation of secondary electron-positron pairs ($e^{\pm}$) interacting with the galactic magnetic field, (2) bremsstrahlung $\gamma$-rays, and (3) and $\gamma$-rays from Compton interactions with photons of interstellar radiation fields~\cite{Buch:2015iya, Cirelli:2010xx}.

There have been various studies of astrophysical $\gamma$-ray observations with the hope of finding $\gamma$-ray signals of dark matter annihilation. These studies included the galactic center region~\cite{Hooper:2011}. However, the interpretation of the observed excess of $\gamma$-rays from the galactic center~\cite{Ajello:2016} is complicated by other possible contributions from sources such as millisecond pulsars. Studies of $\gamma$-ray emission
from dwarf galaxies~\cite{Ackermann:2015} and the nearby galaxies M31 and M33~\cite{DiMauro:2019} have lead to mass-dependent constraints on DM annihilation.

The observed galactic radio spectrum used to estimate the galactic temperature is a result of several measurements. We particularly highlight the measurements performed at frequencies above 60 GHz by COBE/FIRAS instrument, some surveys  at  22, 45, 408 and 1420 MHz \cite{Agrupados} and the measurements of the balloon-borne experiment ARCADE--2, which has measured radio signals  of the sky temperature at frequencies in the range of 3 to 90 GHz.  The ARCADE--2 experiment  detected a significant component of isotropic emission at frequencies between 22 MHz and 10 GHz. This component was brighter than the expected contributions from possible extragalactic sources. This emission is now commonly known as the ARCADE--2 excess \cite{Fixsen:2009xn}. Several mechanisms have been explored over the last decade in an attempt to explain the excess measured by the ARCADE experiment. Although the most accepted explanation is that the excess, due to its isotropy,  is of extragalactic origin, the ARCADE experiment performed measurements in a limited region of the sky. 

Concerning to DM interpretations of such excess, the results given by ARCADE--2 collaboration indicate that the spectral excess was hard, requiring  annihilation models with a hard electron-positron spectrum $dN_{e}/dE$. This behavior can be produced by models with a large decay or annihilation branching ratio into leptons. Among them, many authors considered DM annihilation and decay into leptonic channels such as $\mu^{+},\mu^{-}$ combined with a thermal annihilation cross section $\langle \sigma v\rangle=3\times 10^{-26}$ cm$^{3}$/s in order to agree with the observations \cite{Fornengo:2011cn}.
However, the values previously considered for thermal annihilation cross sections and the mass range are now in tension with the strong bounds on annihilating DM and decaying DM for all channels~\cite{Cirelli:2016mrc}.

Using a $\chi^{2}$ analysis, it was concluded that a light DM candidate annihilating mainly into leptons, not coupling dominantly to quarks, was the best choice in attempting to explain the excess \cite{Fornengo:2011cn}. Similar conclusions held for the decaying DM case \cite{Fornengo:2011cn}.
Other more ``traditional" DM scenarios have been considered with a DM candidate with mass $m_{DM} \sim 100\,{\rm GeV}$ and annihilation into $b \bar{b}$ quark-antiquark pairs. However, the predicted $dN_{e}/dE$ spectrum was too soft to reproduce the ARCADE--2 data, producing a poor $\chi^{2}$ fit. Scenarios involving heavier DM candidates with dominant hadronic annihilation or decay in the final states are strongly constrained by $\gamma$-ray experiments ~\cite{Fornengo:2011cn}.

In this paper, we address our attention to the study of radio waves of synchrotron radiation of DM candidate that annihilates into leptonic channels. We will reanalyse the radio measurements of the sky temperature of the universe, considering the DM annihilation channels into leptons and also considering the channels involving intermediate bosons, ${\rm DM, DM} \rightarrow V,V \rightarrow 4$ leptons, where $V$ denotes some new light boson. Our main goal is to compare the possible signature of this channel with other DM annihilation channels already studied in the literature. The dark matter annihilation channels discussed in this article have  stronger signatures if compared to other annihilation channels discussed in the literature. From this point of view, could the present mechanism account for both the excess signal and its isotropy? Although Galactic emission has a non-trivial spatial morphology, given the sky region covered by ARCADE sensitivity, it would be interesting to determine the radio signature within the range of latitude and longitude covered by the experiment.

The paper is organized as follows. In Section \ref{sec:Flavored} we consider hypothetical DM models and perform  simulations. In Section \ref{sec:Configurations} we present the astrophysical configurations and synchrotron halo functions necessary to perform the simulations. In Section \ref{sec:Radio} we present our results, giving spectra of radio emission for the DM candidates and models considered. In Section \ref{sec:Conclusion} we present our final remarks about the simulations performed here.

 \section{The Dark Matter Model}
\label{sec:Flavored}

The model studied here involving DM annihilation in two light $V$ bosons with the subsequent decay of $V\rightarrow l^{+}l^{-}$ (where $l$ denotes the $e,\mu,\tau$ leptons) is based on the theory presented at \cite{ArkaniHamed:2008qn}. Such class of theories were proposed in attempt to fit the spectra of ATIC and PAMELA which required  a DM candidate to annihilate at a level very above what was expected for a thermal relic. It's known that the typical annihilation into $Z$ bosons won't produce many hard leptons. On the other hand, the annihilation into $W$ boson produces hard leptons, but also many soft leptons via hadronic showers.

\begin{figure*}
\centering
\begin{tabular}{cc}
\includegraphics[width=0.5\textwidth=100mm]{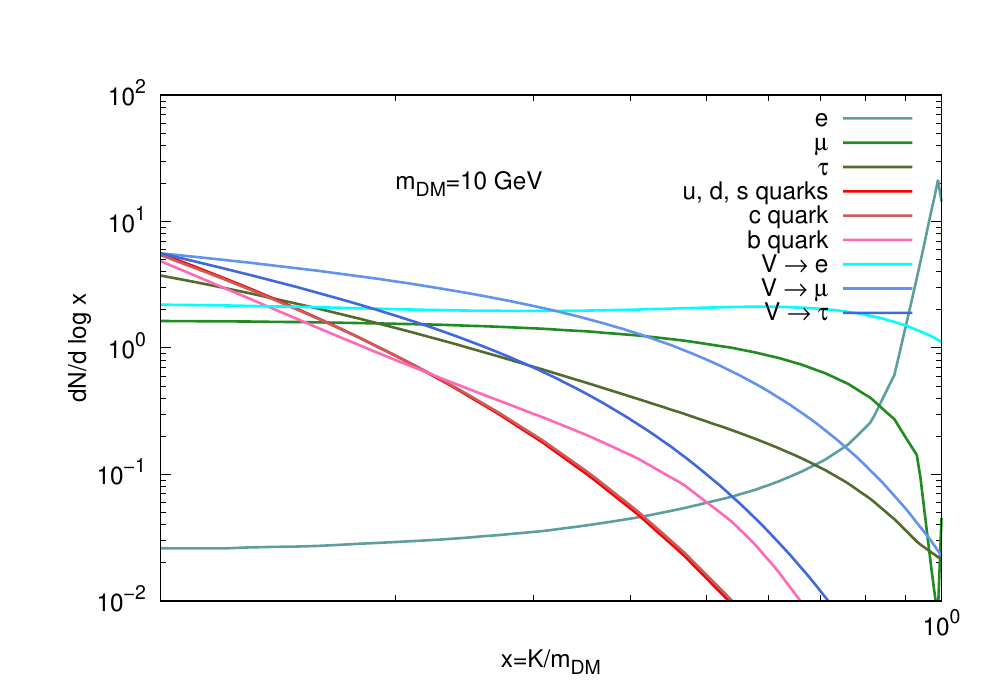}
\includegraphics[width=0.5\textwidth=100mm]{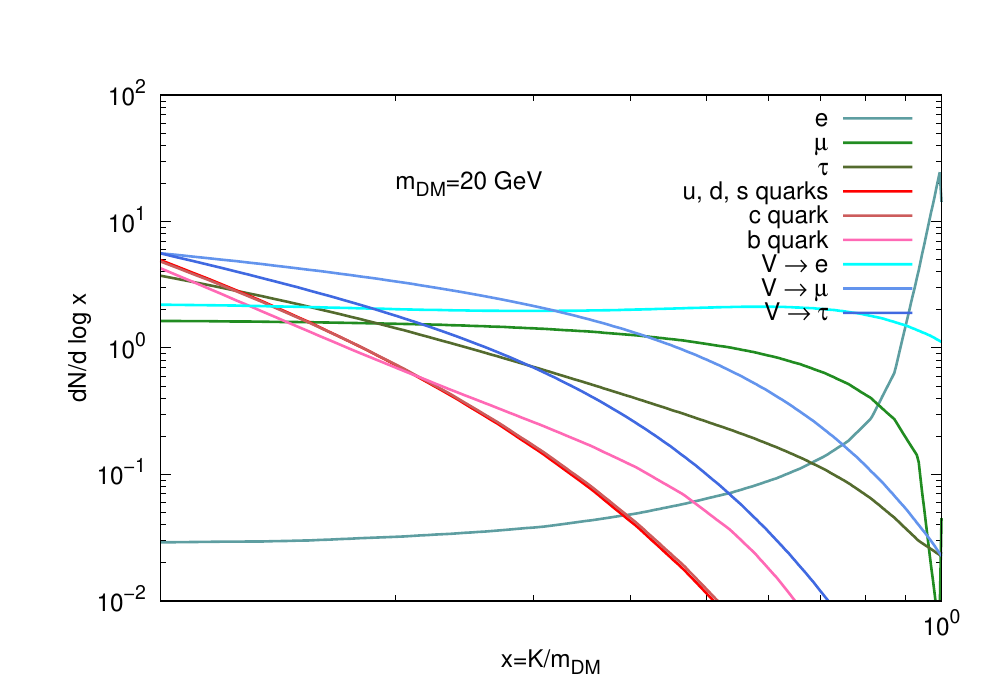}\\
\end{tabular}
\caption{Primary flux of $e^{+}$ spectra of $V$ boson compared with other standard model spectra. The kinetic energy is represented as $K$. Note that the $V\rightarrow e$ channel has a flat behavior within the range $0.1 \lesssim x \lesssim 1$.}
\label{figdNdE}
\end{figure*}

The hadronization and further decay of the primary annihilation products leads to the production of electrons ($e^{\pm}$), neutrinos ($\nu$), protons ($p$), deuterons ($D$) and photons ($\gamma$-rays).  The electron-positron spectrum is represented by ($dN_{e^{\pm}}/dx$), where $x=K/m_{DM}$, and  $K$ denotes the kinetic energy. This spectrum produces a soft cutoff at the kinematical limit $E_{e^{\pm}} = m_{DM}$. The electron-positron spectra of stable products, resulting from a DM annihilation for several channels, are very different. For instance, in Figure \ref{figdNdE}, considering the range $0.1\lesssim x \lesssim 1$ the spectra for the  $u$, $d$, $s$ and $c$ quark channels are all very similar in magnitude and shape. For the other channels the spectral properties are clearly distinct in the plot and we can have hard, soft, slightly softer and soft-hard spectra. The remaining possibilities of analysis consist in considering the heavy quarks and the Higgs boson, but they also produce softer spectra of leptons.

From the model building point of view, we require that even if a model gives a high annihilation yield into leptons, the annihilation yield into hadrons must be low.  This last constraint comes from limits of diffuse galactic $\gamma$-rays, as well as the bounds on $\pi^{0}$ production from DM annihilation from the observed $\gamma$-ray flux from the galactic center. Other hadronic constraints come from the PAMELA experiment. Many theoretical models for  WIMP  annihilation predict a  substantial  branching ratio to antiprotons.  However,  measurements obtained from the AMS-02, BESS, and PAMELA detectors show  no significant excess of antiprotons above those expected from secondary cosmic ray production\cite{Wechakama:2013hra, Kappl:2014hha, Kappl:2015bqa}. Those results thus put severe limits on particle dark matter models, suggesting  that  such  models should annihilate predominantly to lepton-antilepton pairs rather than hadrons.

 In this work, like other beyond standard model extensions, we had  included a dark sector which can interact with the SM sector via a light mediator V. Then,  we studied possibilities of fitting the ARCADE-2 excess through DM annihilation in such mediator which subsequently decays to standard models leptons. We also assume that this new channel is the dominant one for the studies of DM annihilation models involving $V$ boson.

Empirical bounds on the couplings of these $V$-bosons in this type of model exist. These constraints depend on the scale of the physics involving the $V$-boson. High energy colliders are sensitive to $V$-bosons heavier than 10 GeV; constraints for the lighter $V$-bosons are obtained from precision of QED observables, e.g $B$-meson decay involving fixed target experiments. Besides the constraints on the $V$-boson mass, a constraint on the lifetime of $V$-boson also exists. Its lifetime should be less than one second in order to guarantee that the $V$-boson decays before the onset of big-bang nucleosynthesis\cite{Fortes:2015qka, Pospelov:2007mp}.

Cholis et al.~\cite{Cholis:2008vb} first considered the mechanism of DM annihilation into light bosons as an explanation for the excess of cosmic ray positrons observed by HEAT experiment without an accompanying excess of $\gamma$-rays and $\bar{p}$. If the mixing with the standard model is small, the dependence on the decay models comes from the kinematics. This favors the production of electrons, muons, charged pions consistent with the constraints on $\pi^{0}$'s and $\bar{p}$. Other studies for this type of model involving Sommerfeld enhancements from new forces have considered three basic candidates for the $V$ boson: a light scalar, a pseudoscalar $\pi$ with a preferred goldstone like coupling to matter, and a spin -1 boson from some gauge symmetry \cite{ArkaniHamed:2008qp}.

For the simulations considered here, we have worked out a model independent analysis for a dark matter candidate. As particle physics input, we have as free parameters the mass of the candidate and the annihilation cross section. We have directed our analysis to direct  or via mediator DM annihilation into leptonic final states.

In the next paragraphs we briefly explain  how some of the models considered here could fit the experimental radio observation of sky temperature. Figure \ref{figdNdE} shows the $e^{+}$ spectra of $V$ boson in comparison with other standard model spectra, considering, e.g., $m_{DM} = 10$ and $20\,{\rm GeV}$. It is possible to verify that there is no great variation of the spectra within these mass ranges.

${}$

In general, an hypothetical DM candidate from some model can annihilate in the primary channels of SM involving the known particles: $u-$ type quarks, $d-$ type quarks, $e^{\pm}$, $\mu^{\pm}$, $\tau^{\pm}$, $\nu_{e}$, $\nu_{\mu}$, $\nu_{\tau}$,  $W$, $Z$ and $h$ bosons, photon ($\gamma$), gluon ($g$) or can annihilate into a pair of an unknown particle of dark sector not yet discovered.

${}$

The ARCADE--2 results  revealed that the observed radio spectrum is rather hard requiring a hard $e^{\pm}$  spectrum $dN_{e^{\pm}}^{f}/dE$. Considering the range of mass that we are studying here, such a spectrum may be produced by DM annihilation with a large branching ratio into leptons. See Fig 3 of ref \cite{Cirelli:2010xx}. A lighter DM is favored instead of a heavier candidate because the heavier one presents hadronic annihilations and decays having final states that are strongly constrained by the $\gamma$-ray producing channels \cite{Fornengo:2011cn, Elor:2015tva}.

${}$

For comparison purposes, in our analysis we had also considered the annihilation channels $DM,DM \rightarrow e^{+},e^{-}$ and $DM,DM \rightarrow \mu^{+},\mu^{-}$ besides the  annihilation channel $DM,DM \rightarrow V,V\rightarrow 2l^{+},2l^{-}$. This last annihilation process  will occur through s-channel exchange of the mediator into two leptons and two antileptons.

${}$

In this paper we do not analyse the phenomenology of the DM model since it depends sensitively on the couplings of DM to SM particles, the nature of the candidate (Dirac or Majorana Fermion, Scalar), DM interaction with scalars, pseudoscalars, vectors, and axial vectors, its annihilation cross section, etc. The parameter space of such candidate  has to be in agreement with the constraints of both direct and indirect detection experiments. Considering a fermionic DM candidate, depending on its the mass, the direct detection constraints such as LUX can dominate over the indirect detection ones such as the AMS experiment for positrons and the {\it Fermi} $\gamma$-ray telescope.

\section{Astrophysical Configurations and Synchrotron Halo Functions}
\label{sec:Configurations}

In this section, we focus our attention on the synchrotron emission as a secondary radiation of DM. The synchrotron radiation is generated when accelerated ultra-relativistic particles ($e^{\pm}$) interact with electromagnetic fields. The calculations involving such radiation combine several ingredients, with the objective of describing the magnetic field and the functions of propagation of $e^{\pm}$.

Our Galaxy has a complicated magnetic field that can be expressed as the sum of a regular component ($\overrightarrow{B}_{reg}$) and of a turbulent magnetic component ($\overrightarrow{B}_{turb}$).
A necessary step to perform calculations involving this field depends on the choice of a model to the represent it. There are also other astrophysical factors to be considered in describing the propagation of $e^{\pm}$ and in the resulting spectrum of synchrotron radiation. The electrons and positrons that we choose here to consider are those that originate from the annihilation of DM. Thus, it is necessary to choose a profile to describe a density of DM in the Galaxy. Other variables used to estimate the synchrotron radiation involve the propagation function of electron and positron. Taking typical values for the Galaxy, in energies  below $\sim$ 5 GeV, ionization and bremsstrahlung  dominate. At higher energies Compton and synchrotron processes dominate \cite{Floyd}. An attempt to describe the energy loss function in the magnetic fields should take into account the contribution of all these processes. We can write an equation for electrons and positrons propagating in our galaxy as \cite{Stecker:1971,Buch:2015iya}
\begin{equation}\label{cm}
  b_{tot}(E,r,z)\equiv-\frac{dE}{dt}=b_{coul} + b_{ion}+b_{brem}+b_{cs}+b_{syn},
\end{equation}
where $E$ denotes the energy of electron and positron, $r$ and $z$ are cylindrical galactic coordinates and $b_{coul}$ (Coulomb scattering),  $b_{ion}$ (ionization), $b_{brem}$ (bremsstrahlung), $b_{cs}$ (Compton up-scattering) and $b_{syn}$ (synchrotron radiation) denote the ways that the pair $e^{\pm}$ loose energy.

Regarding to Eq. (\ref{cm}), the complete expressions for partial energy losses can be found in references \cite{Buch:2015iya} and \cite{Stecker:1971}.
Now we go into steps to explain  the differential flux of $e^{\pm}$ per unit of energy from DM annihilations or decays in space and time.

 The density of electrons or positrons $f(E,r,z)$ in the position $(r,z)$ is obtained by solving the equation of loss and diffusion \cite{Stecker:1971}, described as
 \begin{equation}\label{pd}
   -\kappa_{0}\left(\frac{E}{GeV}\right)^{\delta}\nabla^{2}f-\frac{\partial}{\partial E}(b(E,r,z)f)=Q(E,r,z).
 \end{equation}
 The first term takes into account the effects of diffusion. Considering the annihilation of DM, which will be the chosen case contemplated in our analysis , $Q$ is given by
\begin{equation}\label{qann}
  Q=\frac{1}{2}\left(\frac{\rho}{m_{\rm DM}}\right)^{2}\sum_{f} BR_{f}\langle \sigma v\rangle\frac{dN_{e^{\pm}}^{f}}{dE},
\end{equation}
where $\rho$ denotes one of the profiles for distribution of DM, $\sum_{f}BR_{f}$ is the branching ratio for the fermion, $m_{\rm DM}$ denotes the mass of DM, $\langle \sigma v\rangle$ is the annihilation cross section of DM, $dN_{e^{\pm}}^{f}/{dE}$ is the convolution of the injection spectrum or electron injection spectrum  from  DM  annihilations  in  a given final state channel $f$. In annihilation processes, it is related to the source term by Eq. \ref{qann}.

The solution of Eq. (\ref{pd}), which involves Green's functions, is presented in Eq. (\ref{solution}) for the case of annihilation process.
 \begin{equation}\label{solution}
  \frac{d\Phi_{e^{\pm}}}{dE}=\frac{c}{4\pi}f(E,r,z)=\frac{c}{4\pi b(E,r,z)}\frac{1}{2}\left(\frac{\rho}{m_{\rm DM}}\right)^2\sum_{f} BR_{f}\langle \sigma v\rangle\int_{E}^{m_{\rm DM}}dE_{s}\frac{dN_{e^{\pm}}^{f}}{dE}(E_{s})\cdot I(E,E_{s},r,z),
 \end{equation}
where $\Phi_{e^{\pm}}$ is the energy spectrum of electrons/positrons, $I(E,E_{s},r,z)$ is the generalized halo functions, which encode the spatial information of the source term \cite{Fornengo:2011iq}. Basically they are  Green’s functions from a source energy $E_{0}$ to the energy $E$, $c$ is the velocity of light and $b(E,r,z)$ is the energy loss function for electrons and positrons in our galaxy.

The synchrotron intensity depends on the results presented in  Eq. (\ref{solution}). For further details of such calculations we direct the reader to references \cite{Buch:2015iya, Cirelli:2016mrc, Stecker:1971}. We are interested in the synchrotron intensity presented in Eq. (\ref{synchrotron}) at an arbitrary frequency $\nu$ and for known galactic coordinates $(\emph{l},\emph{b})$. For the annihilation case, we can express synchrotron intensity as
 \begin{equation}\label{synchrotron}
  J(\nu, \emph{l},\emph{b})=\frac{1}{2}\left(\frac{\rho_{\odot}}{m_{\rm DM}}\right)^2\int_{m_{e}}^{m_{\rm DM}}dE_{s}\sum_{f} BR_{f} \langle \sigma v\rangle\frac{dN_{e^{\pm}}^{f}}{dE}(E_{s})\cdot I_{syn}(E_{s},\nu,\emph{l}, \emph{b}),
 \end{equation}
where $I_{syn}(E_{s},\nu,\emph{l}, \emph{b})$ denotes the generalized synchrotron halo function, which contain information about the synchrotron power, the energy loss according to the energy E and its unities are in erg/Hz \cite{Buch:2015iya}.
For practical purposes in order to reach the radio observations, we need to express the Eq. (\ref{synchrotron}) in terms of brightness temperature $T$ in Kelvin(K).
\begin{equation}\label{temperature}
  T(\nu)=\frac{c^{2}J(\nu)}{2\nu^{2}k_{\rm B}},
   \end{equation}
  where $k_{\rm B}=1.38\times 10^{-16}$ ${\rm erg}\,{\rm K}^{-1}$ is the Boltzmann constant.

\section{Results of Radio Emission from  Dark Matter Candidate}
\label{sec:Radio}

In order to perform the calculation involving temperature as a function of synchrotron radiation we have used the codes Cirelli et al.~(\cite{Cirelliweb}). There is a collection of astrophysical configurations to be chosen in order to perform the simulations, like the DM density profile, the propagation of the pair $e^{\pm}$ in the galaxy and the magnetic configurations. Details describing these subjects are given in Refs.~\cite{Cirelliweb, Buch:2015iya}.

For a DM density profile configuration, we chose the universal NFW profile for cold dark matter given in Ref.~\cite{NFW}. This profile is expressed as
 \begin{equation}\label{DMdensity}
 \rho_{\rm NFW}(r)=\rho_{s}\dfrac{r_{s}}{r}\left(1+\frac{r}{r_{s}} \right)^{-2}
 \end{equation}
where the parameters $\rho_{s}$ and $r_{s}$ are the scale density and scale radius and vary from halo to halo taking the values
 $\rho_{s}=0.184$ GeV/cm$^3$ and $r_{s}=24.42$ kpc, assuming that this profile is  normalized by requiring that other parameters as the total  mass of the Milky Way and the density at the location of the Sun take the values $M_{\rm MW}=4.7\times 10^{11} M_{\odot}$, $\rho_{\odot}=0.3$ GeV/cm$^{3}$ and  $r_{\odot}=8.33$ kpc.

To describe the propagation of $e^{\pm}$, we chose the parameters of model MED, viz., $\delta=0.70$, $\kappa_{0}=0.0112$ kpc$^2$/Myr and L= 4 kpc. For magnetic configurations we had chosen the Model MF3 which corresponds to the choice of $B_{0}=9.5$ $\mu$G, $r_{D}=30$ kpc and $z_{D}=4$ kpc. We note that with this parameter choice for $e^{\pm}$ it is difficult to explain the positron excess detected by PAMELA. In that case, the AMS-02 positrons data favor the MAX-type sets of propagation parameters \cite{Boudaud:2016jvj}.

Considering the particle physics model for DM matter candidate,  we took a range of masses from 5 GeV to 20 GeV in our simulation for leptonic channels $e^{+},e^{-}$ and $\mu^{+},\mu^{-}$, $V,V\rightarrow 2e^{+},2e^{-}$, $V,V\rightarrow 2\mu^{+},2\mu^{-}$ and 10-20 GeV for $V,V\rightarrow 2\tau^{+},2\tau^{-}$.

\subsection{The DM,DM $\rightarrow$ V,V $\rightarrow$ four-lepton Channel}
\label{sec:Resultados}

\begin{figure*}[ht!]
\centering
\begin{tabular}{cc}
\includegraphics[width=0.5\textwidth=100mm]{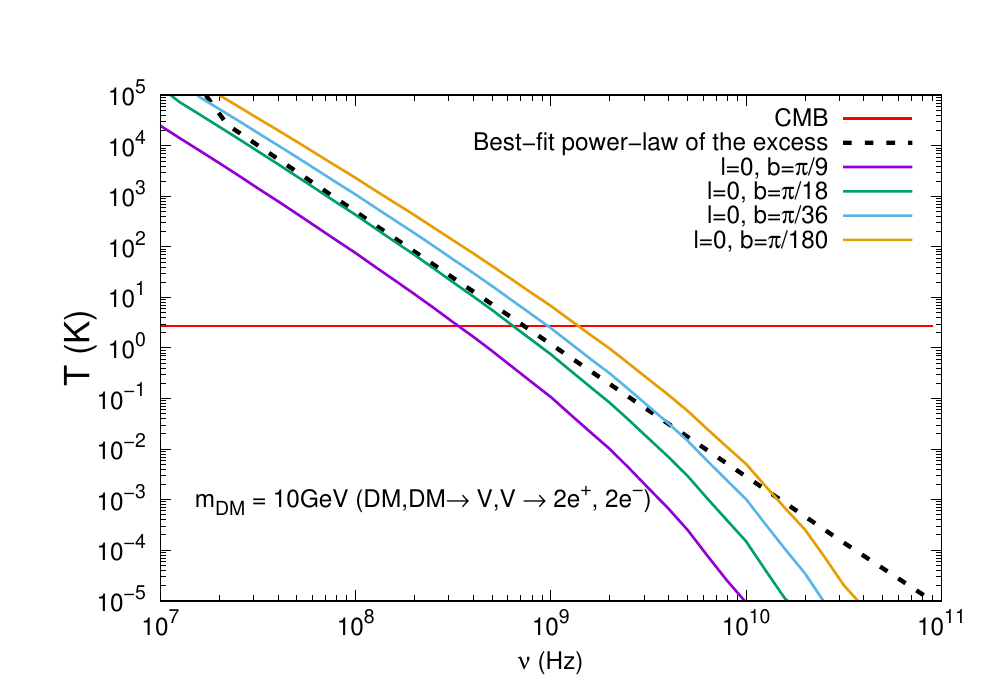}
\includegraphics[width=0.5\textwidth=100mm]{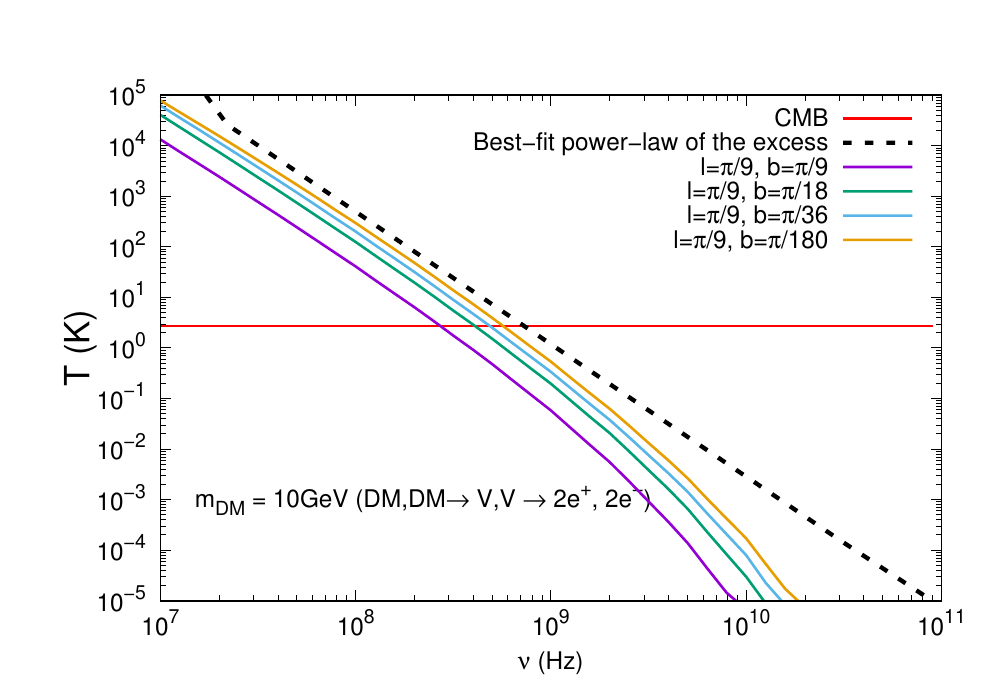}\\
\includegraphics[width=0.5\textwidth=100mm]{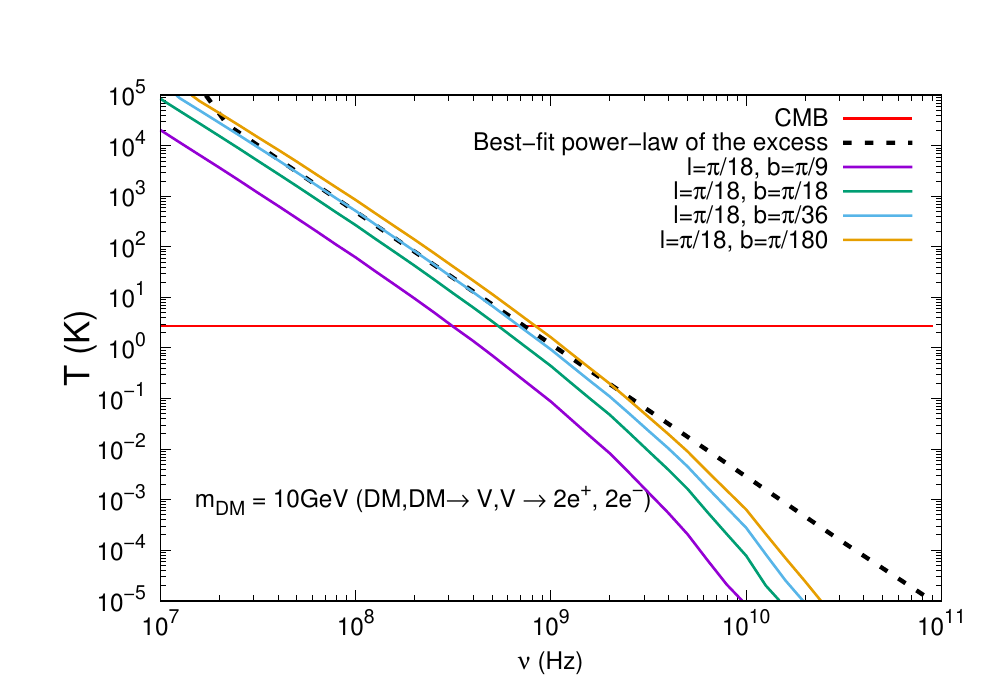}
\includegraphics[width=0.5\textwidth=100mm]{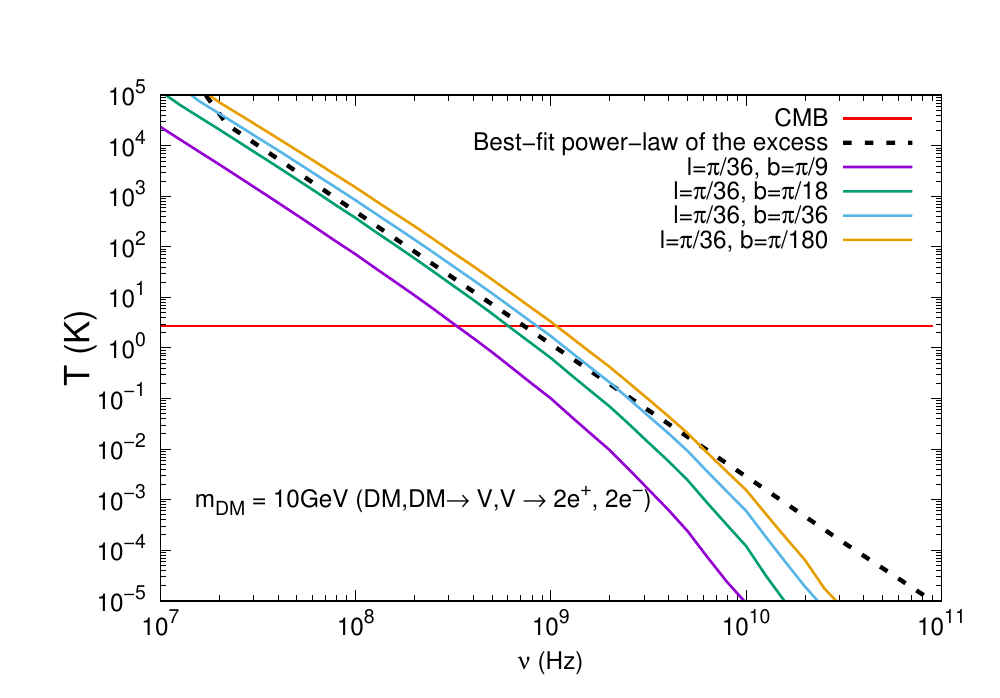}\\
\end{tabular}
\caption{Best--fit power--law of the observed radio excess confronted with the radio signature coming from the annihilation of DM through the channel $DM,DM\rightarrow V,V\rightarrow 2e^{+},2e^{-}$ in a different set of latitude and longitude (in radians). $V$ denotes intermediate vector bosons. The annihilation cross section used for the simulations is  $\langle \sigma v\rangle=3\times 10^{-26}$ cm$^{3}/s$.}
\label{fig:example1}
\end{figure*}

\begin{figure*}[ht!]%
\centering
\begin{tabular}{cc}
\includegraphics[width=0.5\textwidth=100mm]{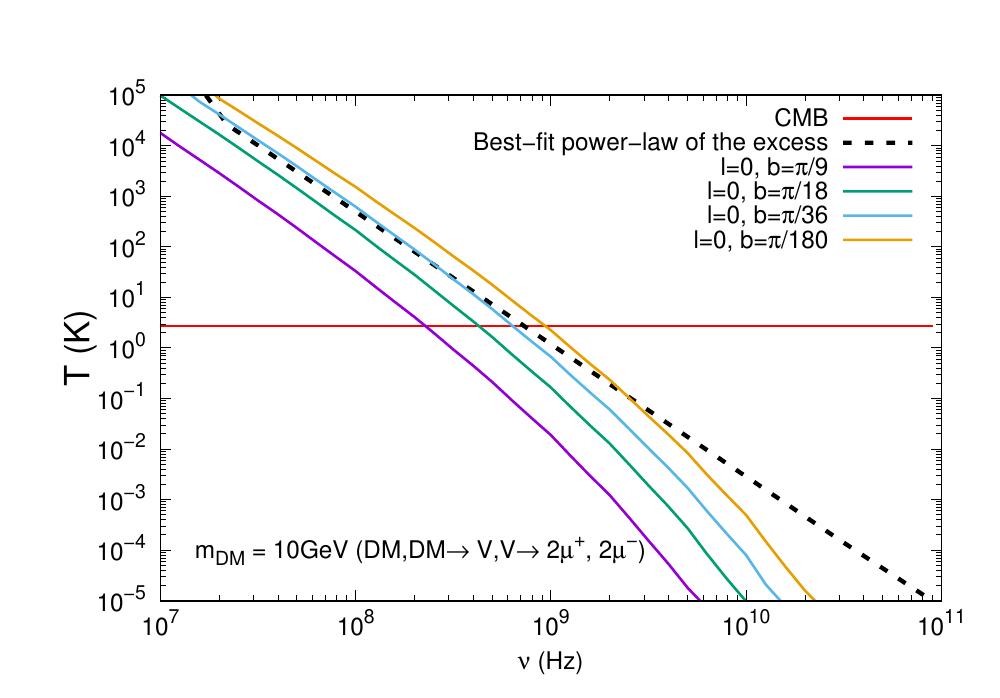}
\includegraphics[width=0.5\textwidth=100mm]{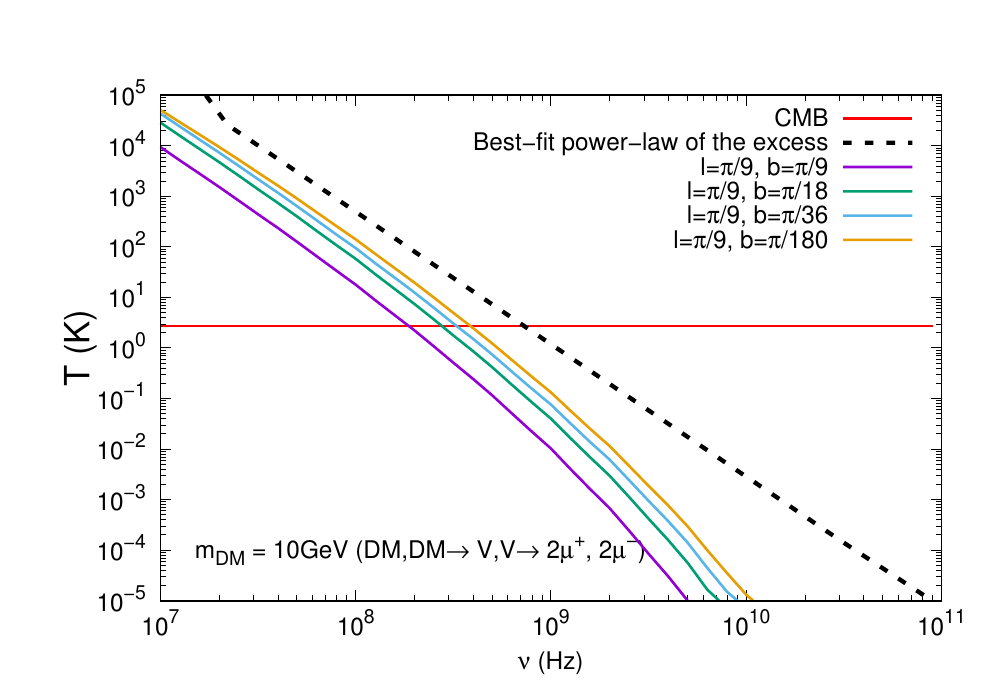}\\
\includegraphics[width=0.5\textwidth=100mm]{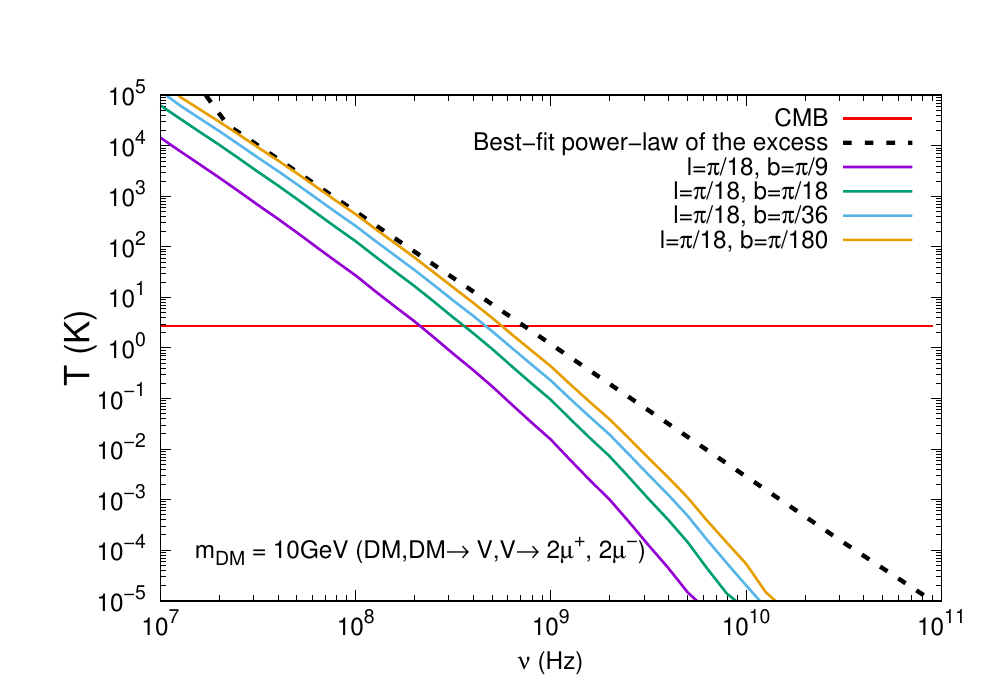}
\includegraphics[width=0.5\textwidth=100mm]{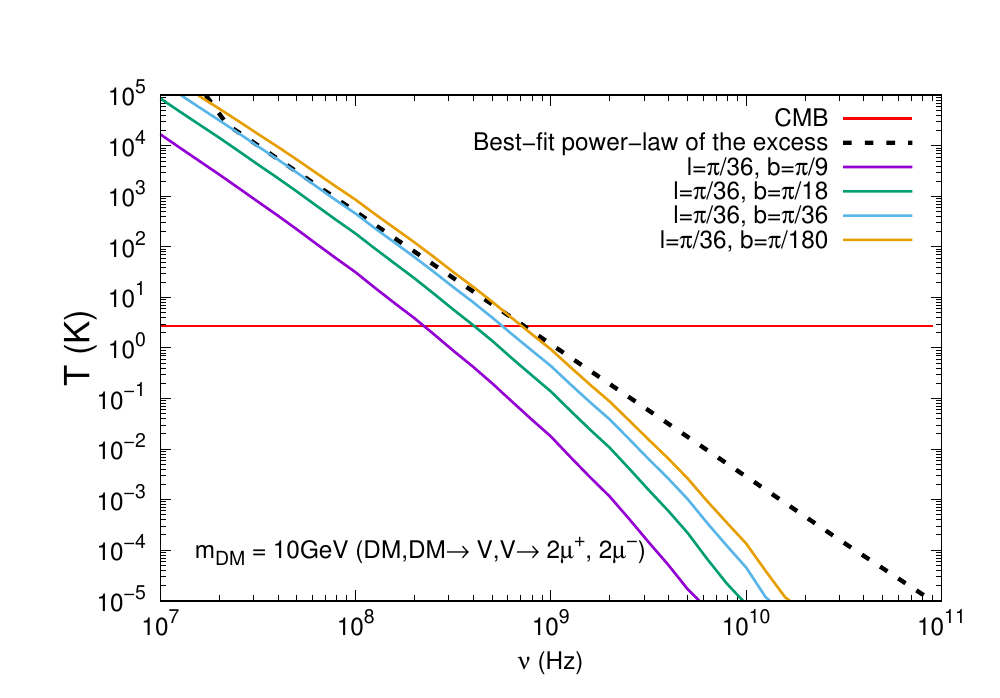}\\
\end{tabular}
\caption{Best--fit power--law of the observed radio excess confronted with the radio signature coming from the annihilation of DM through the channel $DM,DM\rightarrow V,V\rightarrow 2 \mu^{+},2\mu^{-}$ in a different set of latitude and longitude (in radians). The annihilation cross section used for the simulations is  $\langle \sigma v\rangle=3\times 10^{-26}$ cm$^{3}/s$.}
\label{fig:example2}
\end{figure*}

\begin{figure*}[ht!]
\centering
\begin{tabular}{cc}
\includegraphics[width=0.5\textwidth=100mm]{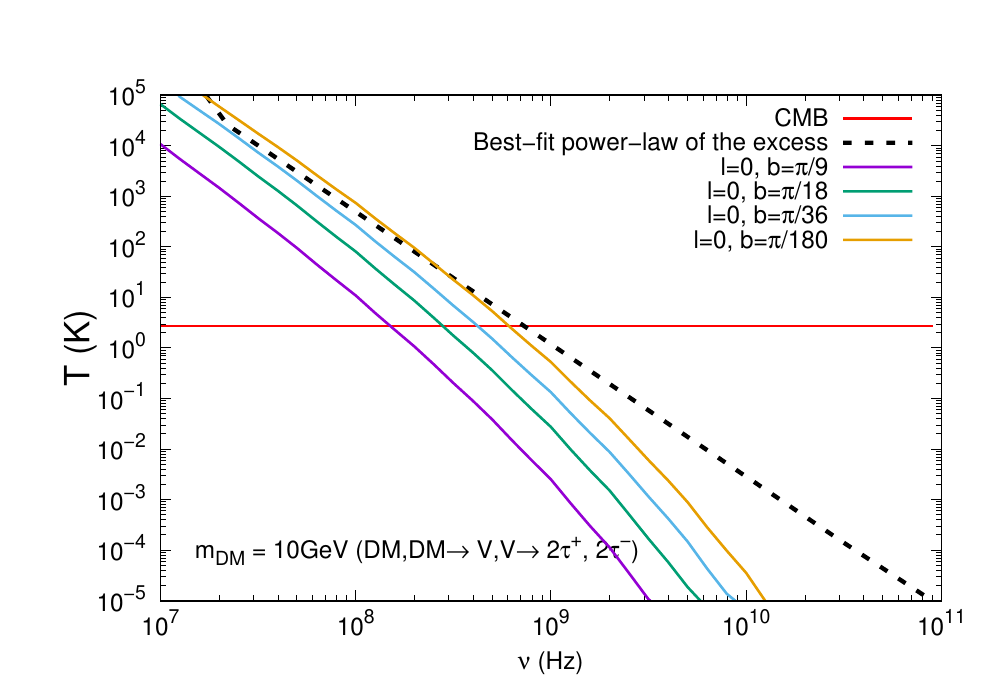}
\includegraphics[width=0.5\textwidth=100mm]{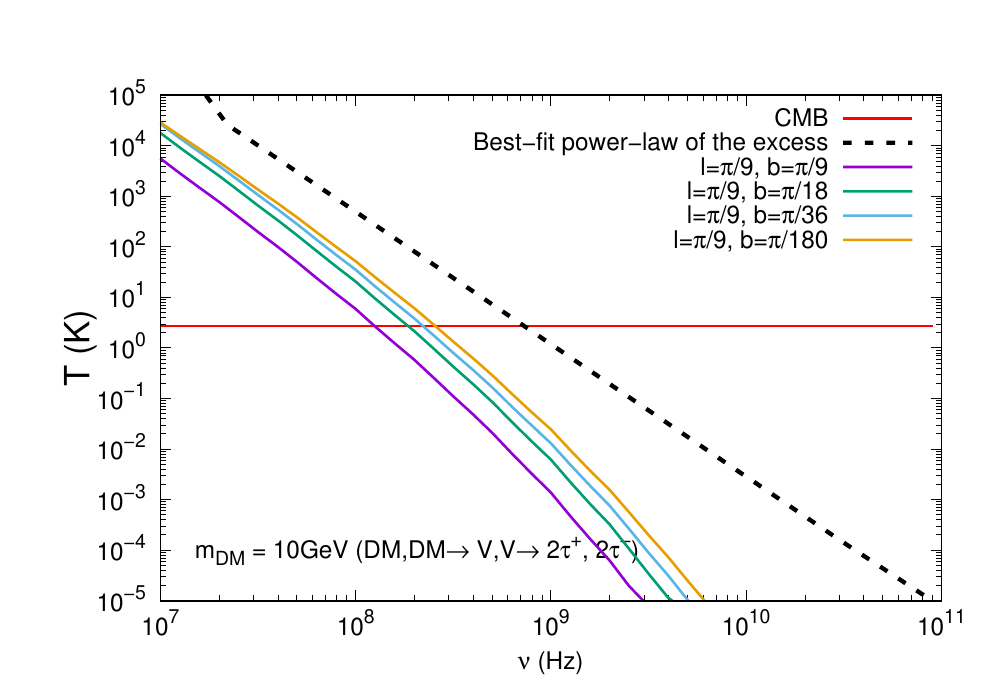}\\
\includegraphics[width=0.5\textwidth=100mm]{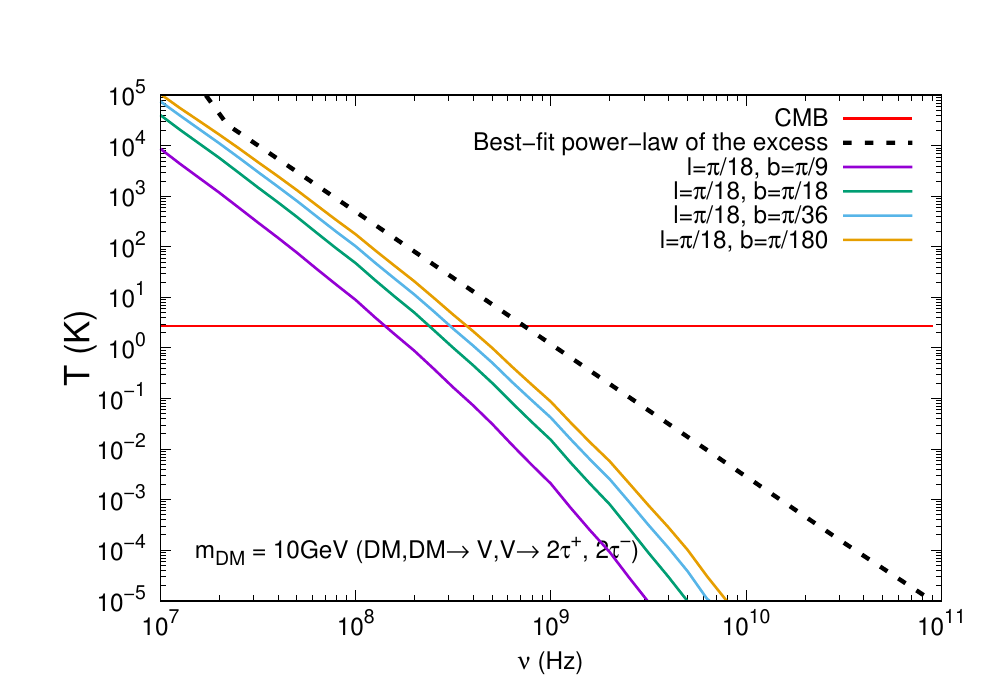}
\includegraphics[width=0.5\textwidth=100mm]{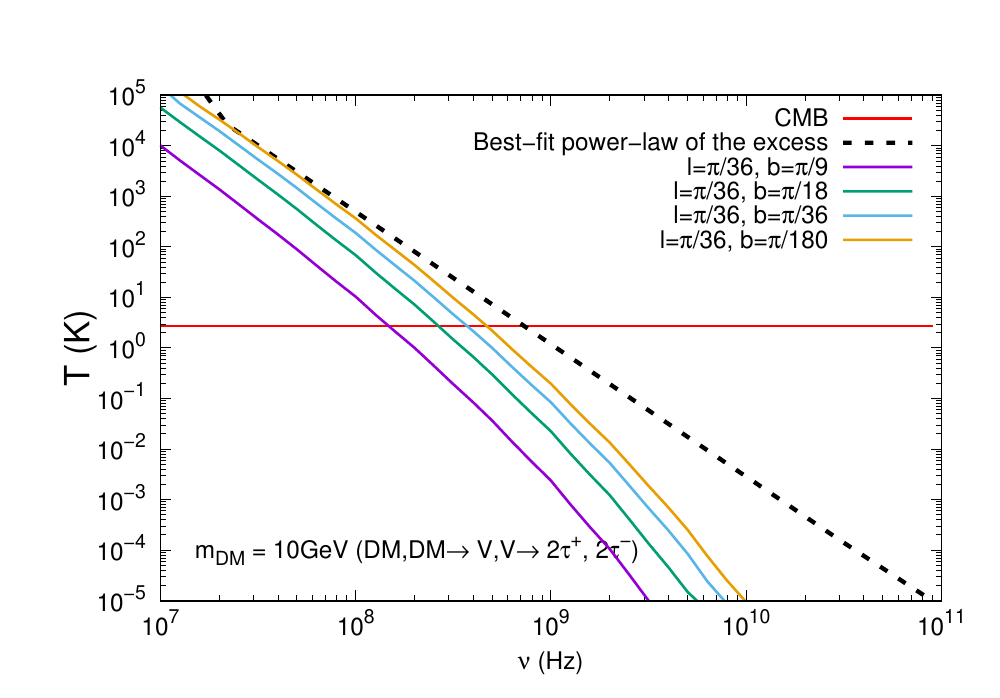}\\
\end{tabular}
\caption{Best--fit power--law of the observed radio excess confronted with the radio signature coming from the annihilation of DM through the channel $DM,DM\rightarrow V,V\rightarrow 2\tau^{+},2\tau^{-}$ in a different set of latitude and longitude (in radians). The annihilation cross section used for the simulations is  $\langle \sigma v\rangle=3\times 10^{-26}$ cm$^{3}/s$.}%
	\label{fig:example3}%
\end{figure*}

In the next graphics we show the results for DM annihilating into a light boson $V$. The studied cases are $V,V \rightarrow 2e^{+},2e^{-}$, $V,V \rightarrow 2\mu^{+},2\mu^{-}$ and $V,V \rightarrow 2\tau^{+},2\tau^{-}$. In order to perform the simulations, we have considered some set of latitude and longitude. For these studied channel, we note that for low latitude and longitude values, the signals generated are very close  to the best fit power-law excess as measured by ARCADE--2 experiment. These channels were not well explored in the literature and their results can be seen at Figures \ref{fig:example1}, \ref{fig:example2} and \ref{fig:example3}.

The combination of low-frequency radio with ARCADE--2 data produced the following fit for the radio excess \cite{Fixsen:2009xn}

\begin{equation}
T(\nu) = T_{R} \left(\frac{\nu}{\nu_{0}}\right)^{\beta},
\label{fixsen}
\end{equation}

\noindent where $T_{R} = 24.1\pm 2.1\,{\rm K}$ is the  normalization for the radio background
being expressed in units of antenna temperature, $\nu_{0} = 310\,{\rm MHz}$ and $\beta = -2.599\pm 0.036$.

The antenna temperature ($T_{A}$) is related to the thermodynamic temperature $T$ by

\begin{equation}
T_{A} = \left( \frac{x}{e^{x}-1}\right) T,
\label{antenna}
\end{equation}

\noindent where $x=h\nu/kT$, $h$ is Planck’s constant, and $k$ is Boltzmann’s constant.

As already noted, for some set of low latitude and longitude it is possible to have a good agreement with the best-fit power law excess, especially for $DM,DM\rightarrow V,V\rightarrow 2e^{+},2e^{-}$ and $DM,DM\rightarrow V,V\rightarrow 2\mu^{+},2\mu^{-}$ channels. The main features of these models are: (i) signals produced by the annihilation of DM fairly reasonably map the radio excess from 20 MHz to about 20 GHz; (ii) the agreement between them is closer to the galactic plane.

Different explanations in terms of astrophysical sources have been explored, but they have not been useful in explaining the measured radio excess. In particular, both the contribution from galactic sources and the usual extragalactic sources have been excluded. The sources of extragalactic origin are mainly discrete radio sources like radio galaxies, although some diffuse sources as synchrotron radiation from clusters and the intergalactic medium have also been explored with no success. Because the origin of the excess remains hidden, more exotic explanations have been proposed. One possibility is that its origin be related to the dark matter.

In principle, the DM annihilation model involving annihilation into vector bosons, producing final state leptons can explain the radio excess. For the curves presented in the Figures (\ref{fig:example1})--(\ref{fig:example3}), we can see that the signature produced by this annihilation channel fits the radio excess up to $b\sim  10$ degrees above the galactic plan and with $l \lesssim 20$.

In order to discuss whether the low-frequency radio experiments in combination with ARCADE--2 could or could not have measured the $DM,DM\rightarrow V,V\rightarrow l^{+},l^{-}$ channel, we will discuss in the next section what ARCADE--2 measured.

\subsection{The ARCADE--2 Experiment}
\label{sec:Arc2}

The second generation of the Absolute Radiometer for Cosmology, Astrophysics, and Diffuse Emission (ARCADE--2) was launched in 2006 and it was designed to measure radio signals of the sky temperature at frequencies in the range of 3 to 90 GHz. In 2009, the results of these measurements (see, e.g., \cite{Fixsen:2009xn}) made available a number of  maps used in the sky coverage analysis. These maps were produced at the central frequencies of 3.3; 8.2 and 10 GHz. ARCADE's final results reported a significant detection of a residual signal that could not be  explained by the CMB plus the integrated radio emission from galaxies estimated from existing surveys. In a recent paper \cite{Singal:2017}, it was stated that there was no clear explanation for this excess emission and that it is  \textit{``by far the least well understood photon background at present....} and, if confirmed {\it it represents a major outstanding question in astrophysics''}.

The ARCADE--2 measurement of CMB temperature is in excellent agreement with the FIRAS measurement at higher frequencies (above 100 GHz). However, at present the origin of such radio excess at frequencies below 3 GHz is unclear, due to the difficult task of modelling the intensity and all-sky spectrum of this signal.

There are other  Galactic and extragalactic interpretations for this excess, such as radio supernovae, radio quiet quasars and diffuse emission from intergalactic medium and clusters  \cite{Fang}, star-forming galaxies, radio galaxies, blazars and millisecond pulsars \cite{Elor:2015tva}.  Many of the above possibilities, despite looking promising, suffer from several constraints coming from multiwavelength observations. In the end, none of them could significantly explain the observed excess.

\begin{figure*}[ht!]
\centering
\begin{tabular}{cc}
\includegraphics[width=0.5\linewidth]{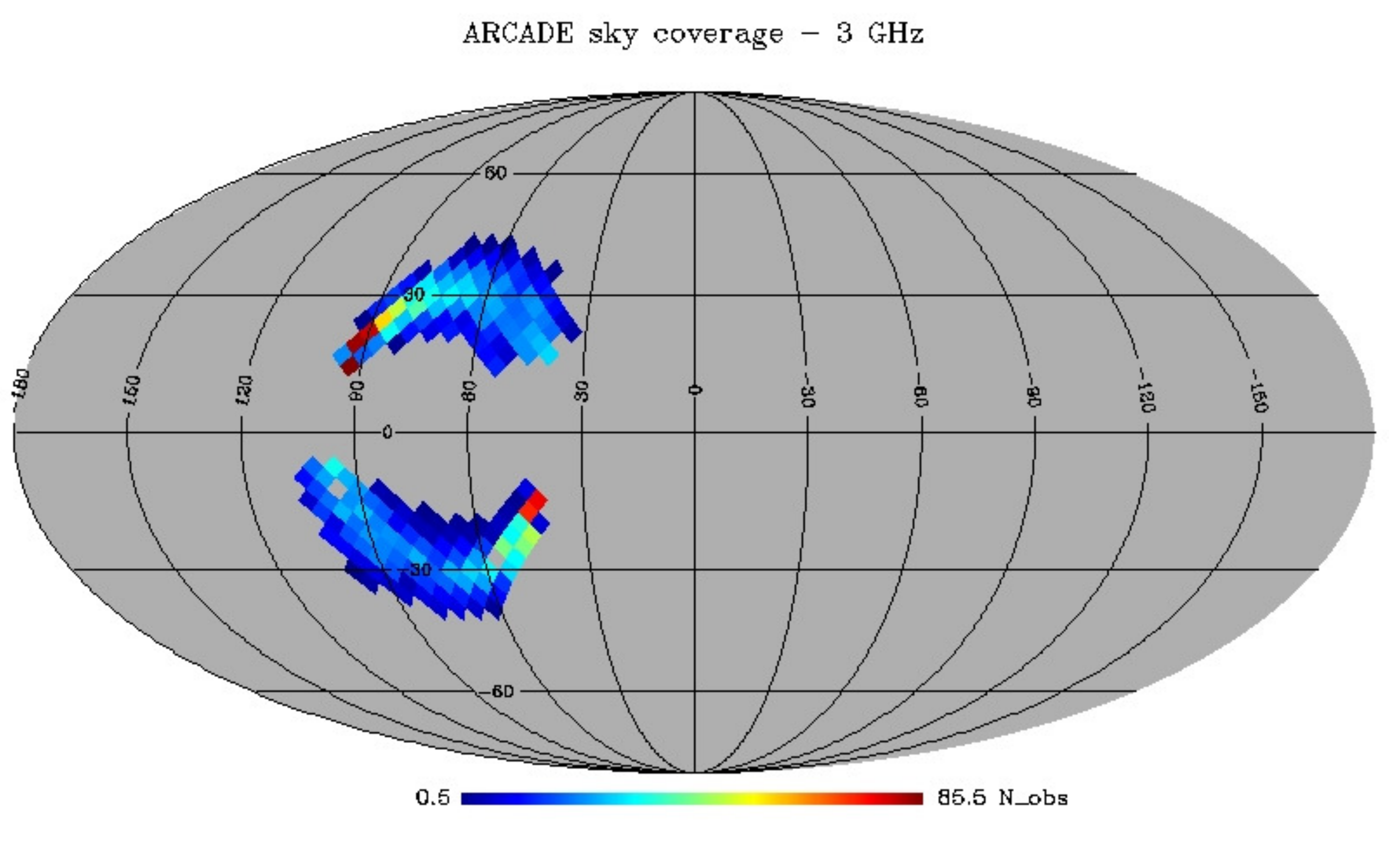}
\includegraphics[width=0.5\linewidth]{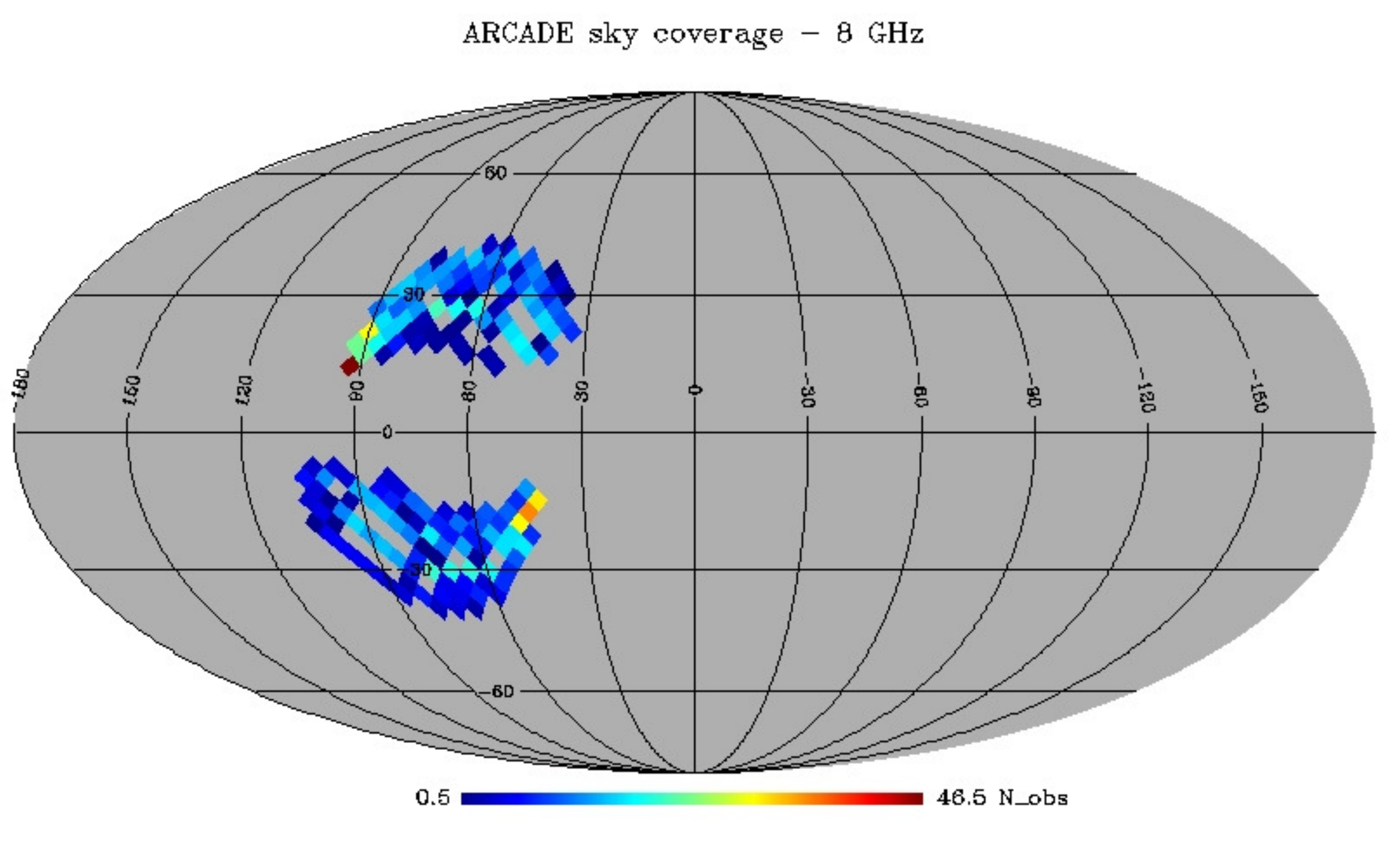}\\
\includegraphics[width=0.5\linewidth]{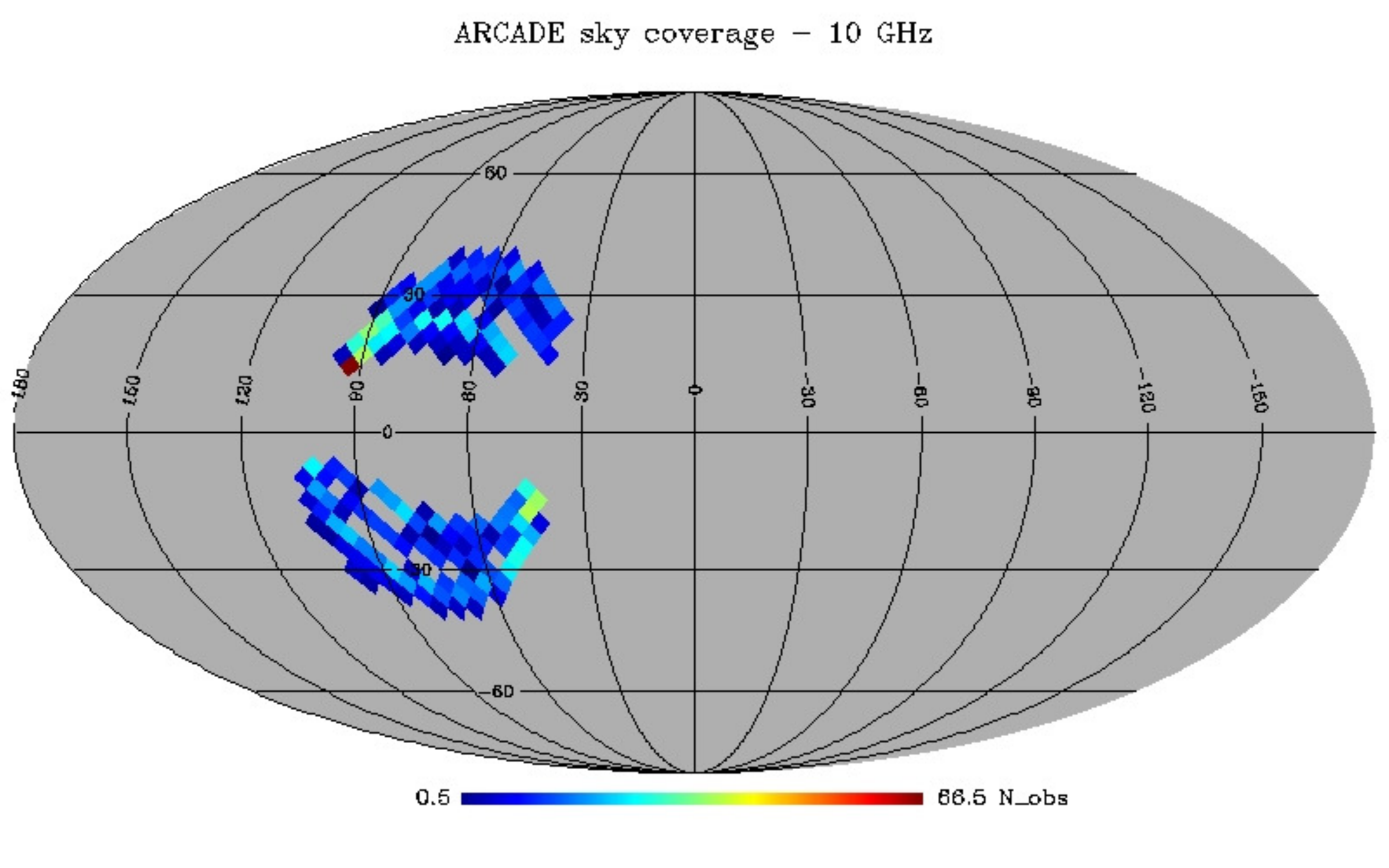}
\end{tabular}
\caption{The panels show the sky area covered by ARCADE--2 measurements. The color palette highlights the number of observations made in a given galactic coordinate as a function of channel frequency. Upper left panel corresponds to the channel 3 GHz, upper right panel to the channel 8 GHz and bottom panel corresponds to 10 GHz.}
	\label{fig1}
\end{figure*}

The sky coverage analysis used in this work uses maps with the number of observations per pixel, which are considered as a sensitivity map. This kind of analysis was not published by the ARCADE 2 team. We had computed the pixel hitmaps considering the number of times each pixel, at each of the ARCADE frequencies (3 GHz, 8 GHz and 10 GHz), was visited during the ARCADE 2 observation time. The panels in Figure \ref {fig1} refer to the area of the sky covered at the above mentioned frequencies. The observed areas in the three ARCADE frequencies cover, approximately, \textbf{$-45^{\circ} \leq {l} \leq +45^{\circ}$} (Galactic latitude) and \textbf\textbf{$+30^{\circ} \leq {b} \leq +120^{\circ}$} (Galactic Longitude, both in J2000 coordinates). However, to pinpoint the most interesting regions in the sky coverage, the values chosen for our analysis are  \textbf{${l} = +15^{\circ}; +30^{\circ}; +45^{\circ}$} and \textbf{${b} = +30^{\circ}; +60^{\circ}; +90^{\circ}; +120^{\circ}$}.

\begin{table}[ht!]
	\begin{center}
	\caption{Counts at given latitude}
	\centering
	\label{table1}
	\begin{tabular}{ccccccc}
		\hline
		\vspace{0.1cm}
		    		&\textbf{3 GHz}  &  			&  			&  			& 	&\\
		\hline
		\vspace{0.1cm}
		\textbf{Latitude/Longitude} & \textbf{30}     & \textbf{45}       & \textbf{60}   & \textbf{} & \textbf{90}     & \textbf{120}\\
		\hline
		\vspace{0.10cm}
		\textbf{15} & \textit{NO} & \textit{23}       & \textit{ 4}   & \textit{} & \textit{85,5}   & \textit{NO}   \\ \vspace{0.10cm}
		\textbf{30} & \textit{NO} & \textit{23}       & \textit{26}   & \textit{} & \textit{16}     & \textit{NO}   \\ \vspace{0.10cm}
		\textbf{35} & \textit{NO} & \textit{10}       & \textit{15}   & \textit{} & \textit{NO} & \textit{NO}   \\ \vspace{0.10cm}
		\textbf{40} & \textit{NO} & \textit{NO}   & \textit{ 1}   & \textit{} & \textit{NO} & \textit{NO}   \\ \vspace{0.10cm}
		\textbf{45} & \textit{NO} & \textit{NO}   & \textit{ 1}   & \textit{} & \textit{NO} & \textit{NO}   \\		
		\\
		
		\hline 		
		\vspace{0.1cm}
				    & \textbf{8 GHz}  &  &  &  & & \\
		\hline
		\vspace{0.1cm}
		\textbf{Latitude/Longitude} & \textbf{30}     & \textbf{45}       & \textbf{60}   & \textbf{} & \textbf{90}   & \textbf{120}\\
		\hline
		\vspace{0.10cm}
		\textbf{15} & \textit{NO} & \textit{NO}   & \textit{NO} & \textit{} & \textit{NO}   & \textit{NO}   \\ \vspace{0.10cm}
		\textbf{30} & \textit{NO} & \textit{13}       & \textit{1}      & \textit{} & \textit{11}     & \textit{NO}   \\ \vspace{0.10cm}
		\textbf{35} & \textit{NO} & \textit{10}       & \textit{15}     & \textit{} & \textit{NO} & \textit{NO}   \\ \vspace{0.10cm}
		\textbf{40} & \textit{NO} & \textit{NO}   & \textit{ 4}     & \textit{} & \textit{NO} & \textit{NO}   \\ \vspace{0.10cm}
		\textbf{45} & \textit{NO} & \textit{NO}   & \textit{NO} & \textit{} & \textit{NO} & \textit{NO}   \\
		\\
		
		\hline
		    &\textbf{10 GHz}  &  &  & &  & \\
		\hline
		\vspace{0.1cm}
		\textbf{Latitude/Longitude} & \textbf{30}     & \textbf{45}       & \textbf{60}   & \textbf{} & \textbf{90}     & \textbf{120}\\
		\hline
		\vspace{0.10cm}
		\textbf{15} & \textit{NO} & \textit{NO}   & \textit{ 4} 	& \textit{} & \textit{66.5}  	& \textit{NO}   \\ \vspace{0.10cm}
		\textbf{30} & \textit{NO} & \textit{ 4}      & \textit{13}    & \textit{} 	& \textit{ 1}     	& \textit{NO}   \\ \vspace{0.10cm}
		\textbf{35} & \textit{NO} & \textit{16}     & \textit{13}    & \textit{} 	& \textit{NO} 	& \textit{NO}   \\ \vspace{0.10cm}
		\textbf{40} & \textit{NO} & \textit{NO}   & \textit{11}     & \textit{} 	& \textit{NO}	& \textit{NO}   \\ \vspace{0.10cm}
		\textbf{45} & \textit{NO} & \textit{NO}   & \textit{ 1}      & \textit{} 	& \textit{NO} 	& \textit{NO}   \\		
		\hline
		\hline
	\end{tabular}
	\end{center}
	{\bf Note.} The acronym NO stands for ``no observations" in that channel.
\end{table}

Table \ref{table1} presents the counts at a given latitude.

\begin{figure*}[ht!]
\centering
\begin{tabular}{cc}
\includegraphics[width=0.45\textwidth=100mm]{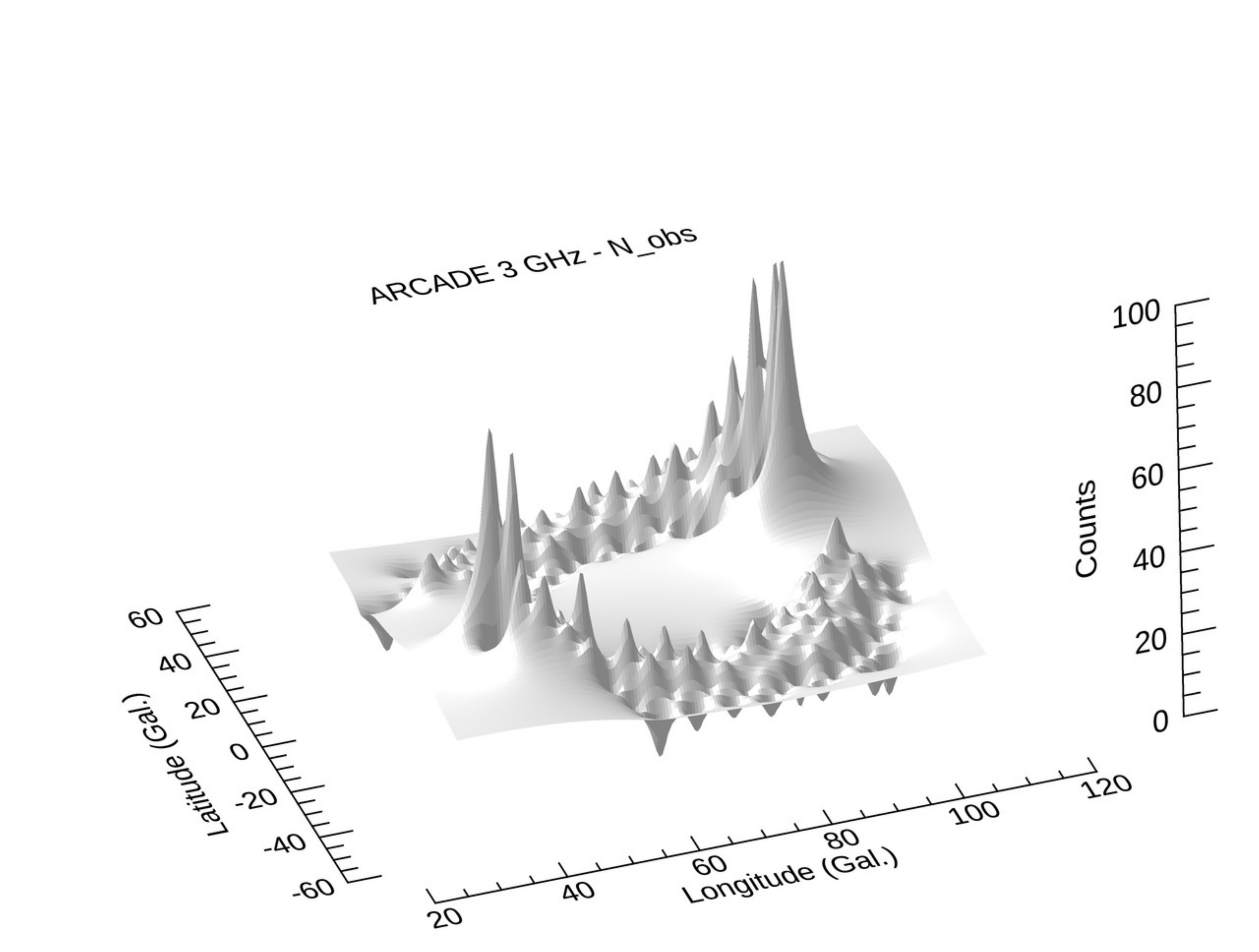}
\includegraphics[width=0.39\textwidth=100mm]{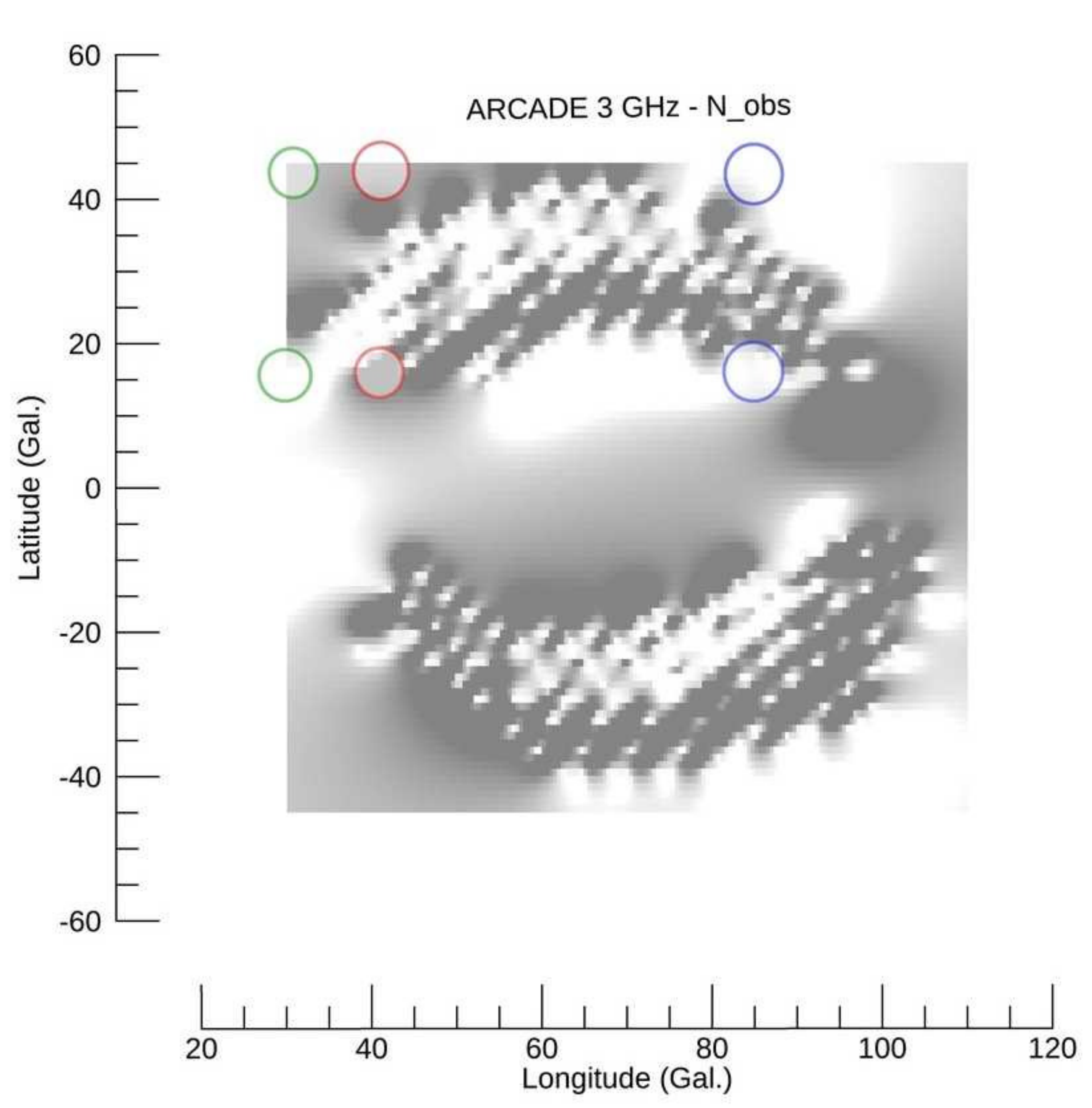}\\
\includegraphics[width=0.45\textwidth=100mm]{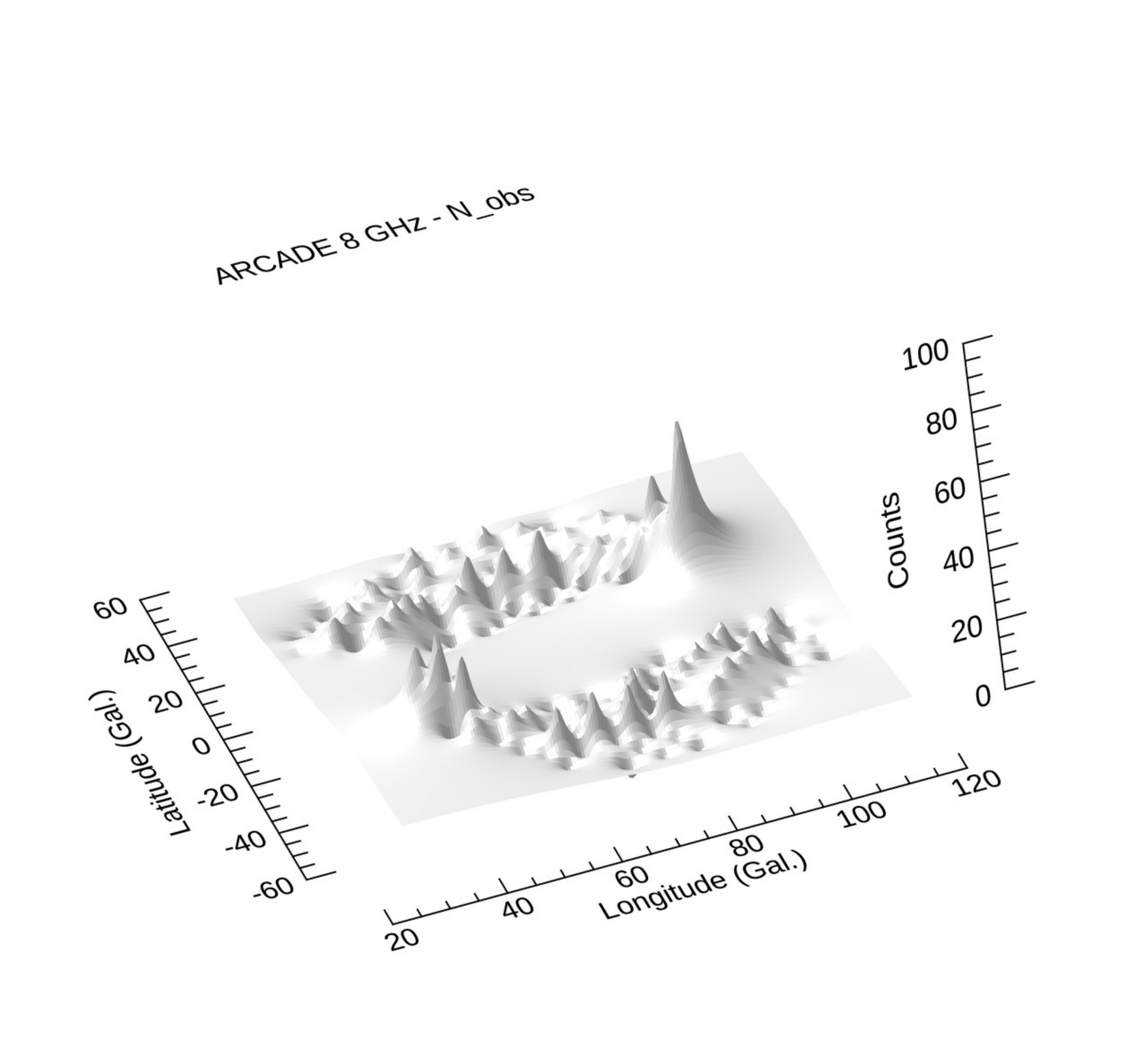}
\includegraphics[width=0.39\textwidth=100mm]{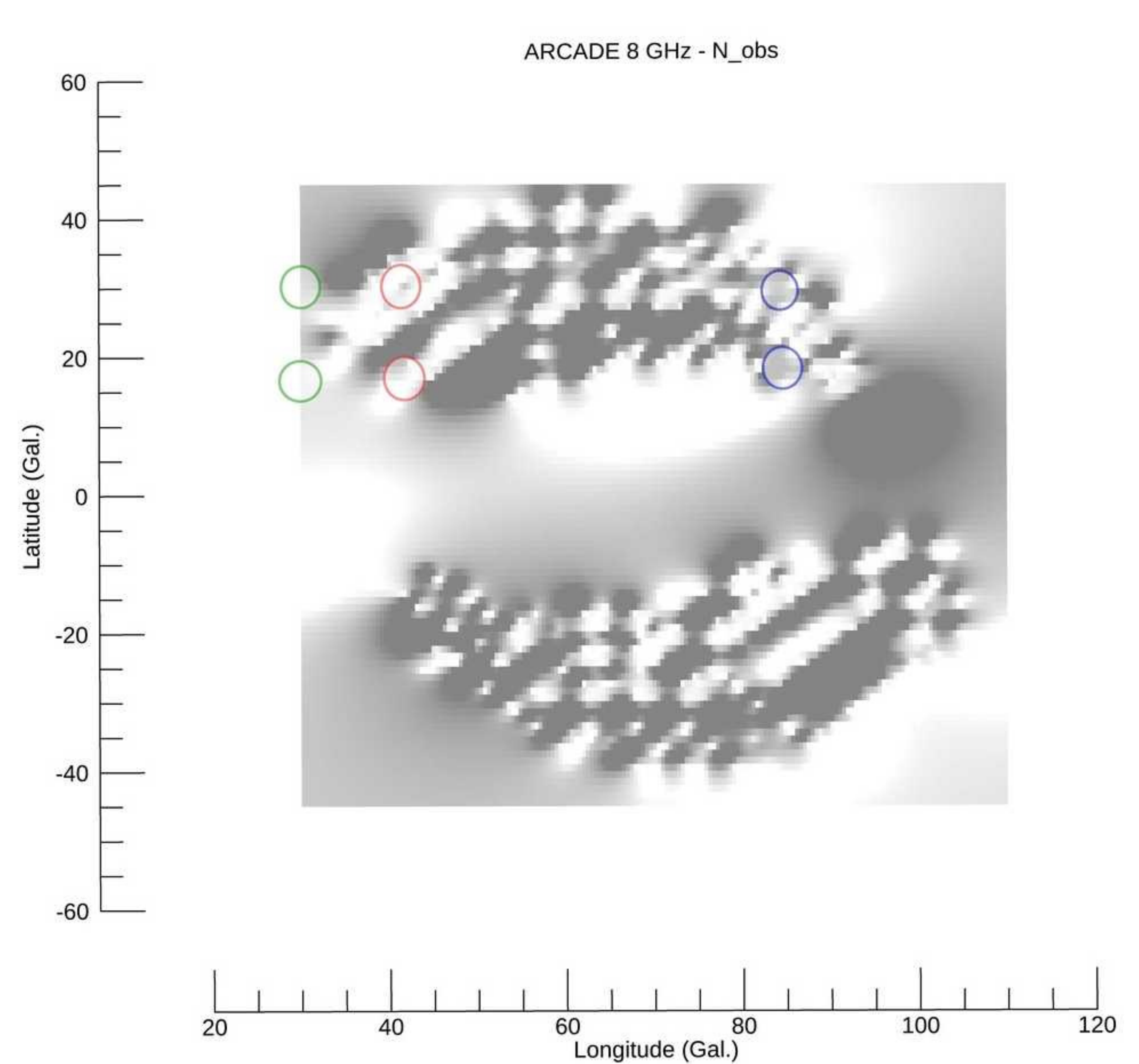}\\
\includegraphics[width=0.45\textwidth=100mm]{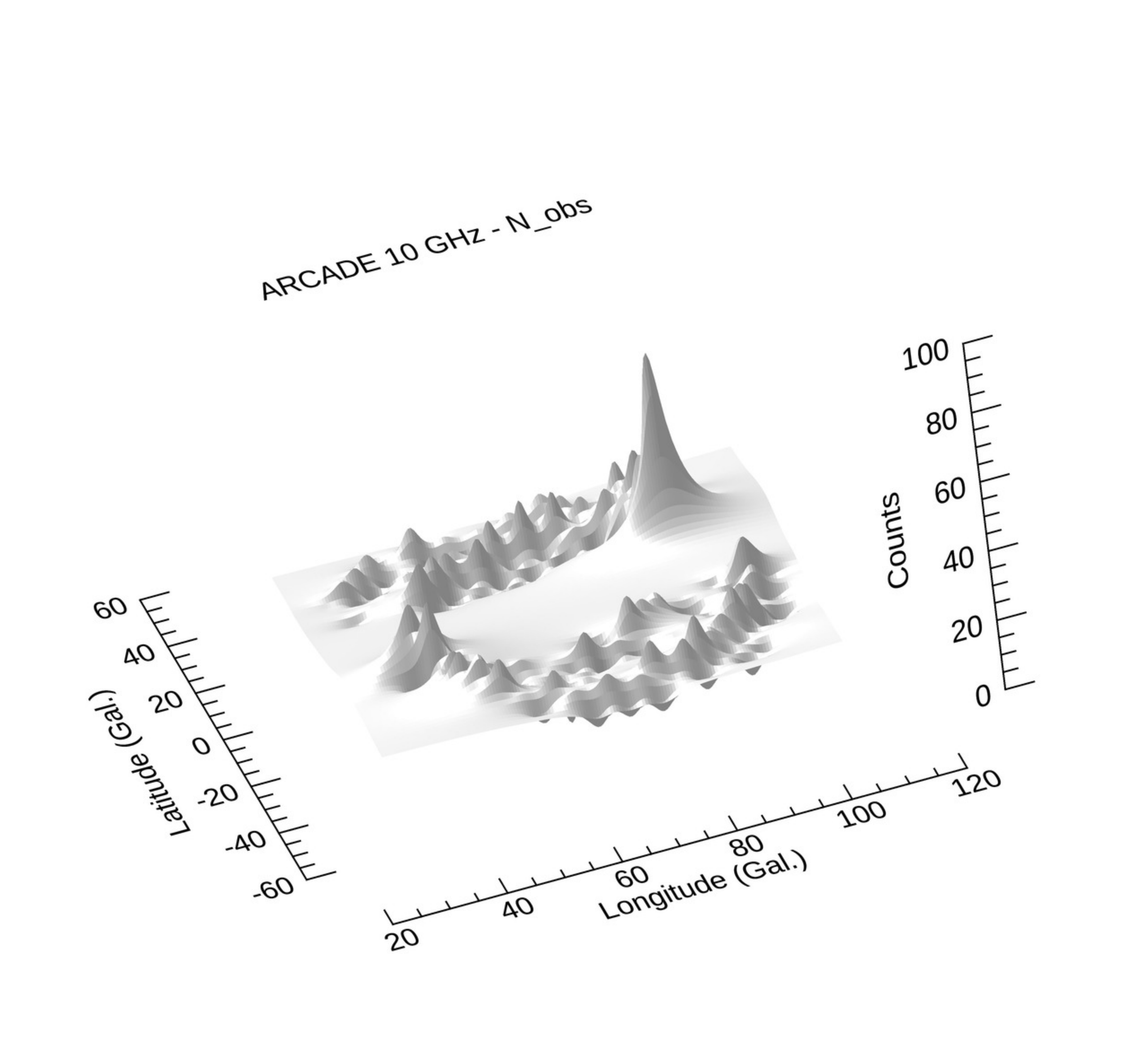}
\includegraphics[width=0.39\textwidth=100mm]{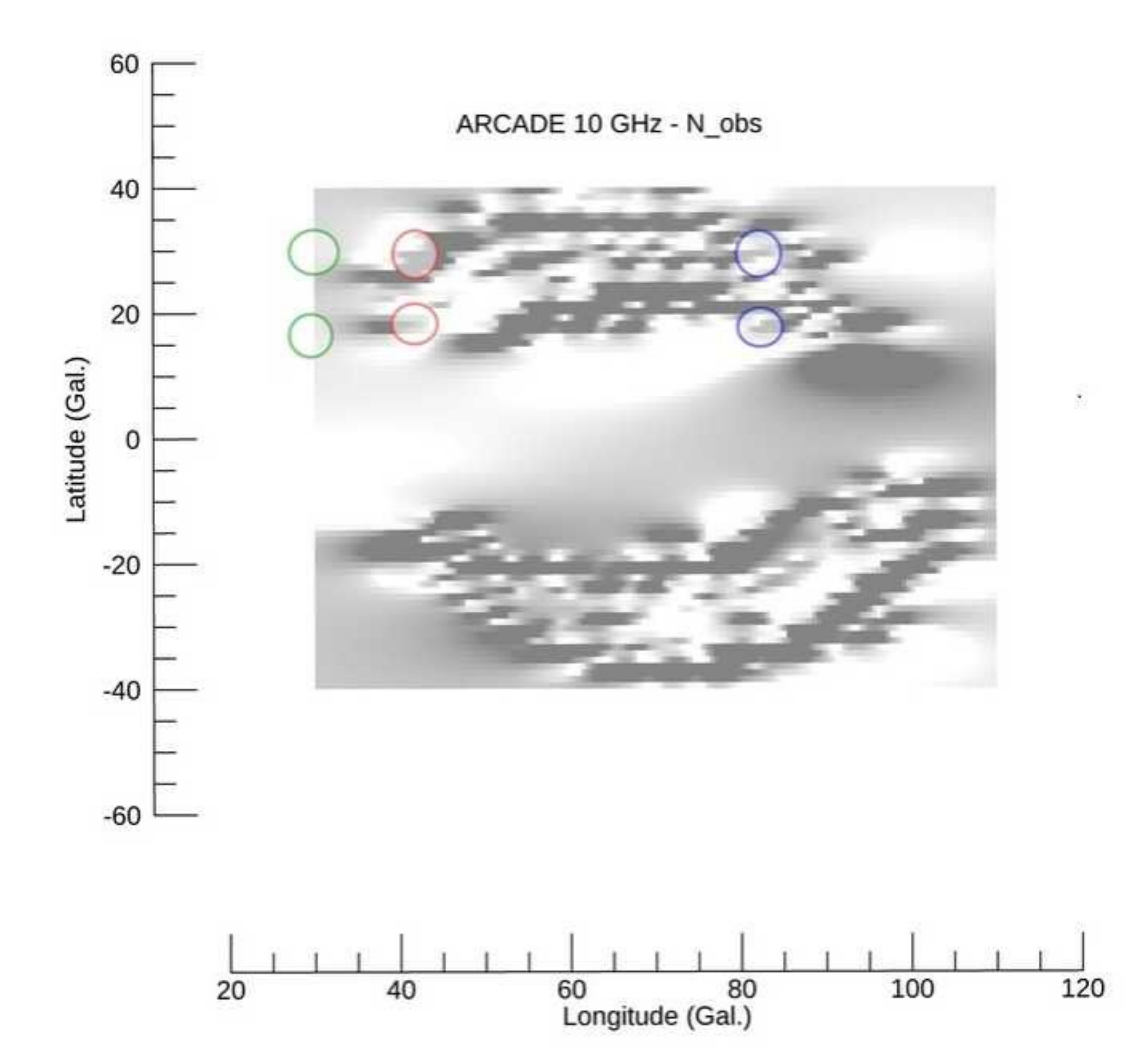}\\
\end{tabular}
	\caption{The panels show the ARCADE--2 observations for the central frequencies of 3GHz (upper left and right panels), 8GHz (middle left and right panels) and 10GHz (bottom left and right panels). The panels on the left side of the Figure correspond to the counting maps while the right side panels (in 2D projection) show the counts in face-on projection. The circulated points green, red and blue denotes respectively the (latitude, longitude) in the following order: for 3 GHz: upper part(30,30); (30,45); (30,85) / lower part (15,30), (18,45) and (15,85);
	for 8 GHz: upper part(30,30); (30,45); (30,83) / lower part (17,30), (17,45) and (18,83);
	for 10 GHz: upper part(30,30); (30,45); (30,83) / lower part (17,30), (19,45) and (18,83).}
	\label{fig4}
	\end{figure*}

Table \ref{table2} contains statistical information for each contour map, viz., the average, maximum, minimum,  variance, standard and absolute deviation and  skewness (see the panels of the Figure \ref{fig4}). The mean values of the standard deviation and absolute deviation differ by a small difference at $ 1 \sigma$ uncertainty level. All of these quantities are very used in statistics, The  skewness measures the asymmetry of the probability distribution of a variable about its average, being positive, negative or even undefined. For the set of latitude/longitude studied here, we have all positive skews, which indicates that the tail is on the right of the distribution.

 The points of greatest diagonal intensity presented in contour maps are signals from the Galaxy that were not cut by the least restrictive mask produced with Planck satellite data (https://lambda.gsfc.nasa.gov/product/planck/).

The flat surface that functions as a ``2D spline" may give the idea that there are no values below it, but the ``face-on" plots in the Figure \ref {fig4} show ``valleys" that correspond to values below this ``spline". Due to the observation strategy and beam size at different frequencies, there are  pixels that were not covered, even being at the border of the scanning area.

\begin{table}[ht!]
	\centering
	\scriptsize
	\caption{Contour Map Data}
	\label{table2}
	\begin{tabular}{cccc}
		\hline
		\hline \vspace{0.10cm}
		\textbf{Variables} & \textbf{3 GHz} & \textbf{8 GHz} & \textbf{10 GHz}  \\
		
		\hline \vspace{0.10cm}
		\textit{Average}              		& \textit{18.5}        & \textit{9.6}       & \textit{12.6}  \\\vspace{0.10cm}
		\textit{Maximum}            		& \textit{85.5}      & \textit{46.5}      & \textit{66.5}  \\\vspace{0.10cm}
		\textit{Minimum}             		& \textit{0.5}     & \textit{0.5}     & \textit{0.5}  \\\vspace{0.10cm}
		\textit{Variance}          		& \textit{226.8}      & \textit{52.7}      & \textit{83.7}  \\\vspace{0.10cm}
		\textit{Std dev}            		& \textit{15.1}      & \textit{7.3}      & \textit{9.1}  \\\vspace{0.10cm}
		\textit{Abs dev}            		& \textit{10.4}      & \textit{5.6}      & \textit{6.8}  \\\vspace{0.10cm}
		\textit{Skewness}           		& \textit{2.1}       	& \textit{1.5}      & \textit{1.8}  \\\vspace{0.10cm}
		\\
		\hline
		\hline
	\end{tabular}
	\end{table}

\subsection{ Reanalysis of the  DM,DM$\rightarrow V,V\rightarrow$  four-lepton channels}
\label{sec:Resultados_2}

\begin{figure*}[ht!]
\centering
\begin{tabular}{cc}
\includegraphics[width=0.5\textwidth=100mm]{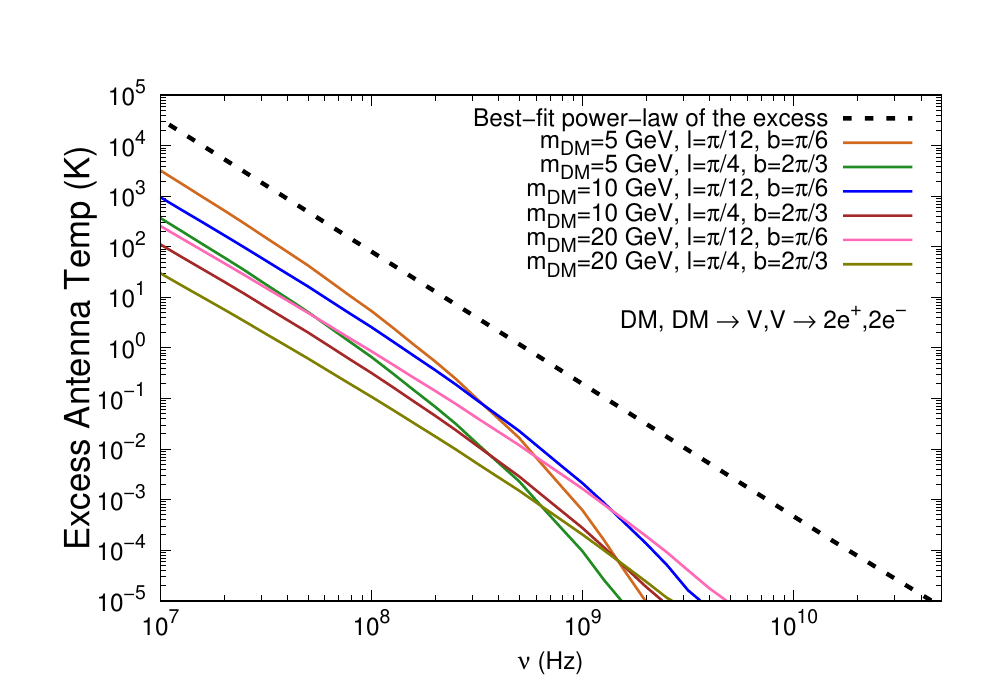}
\includegraphics[width=0.5\textwidth=100mm]{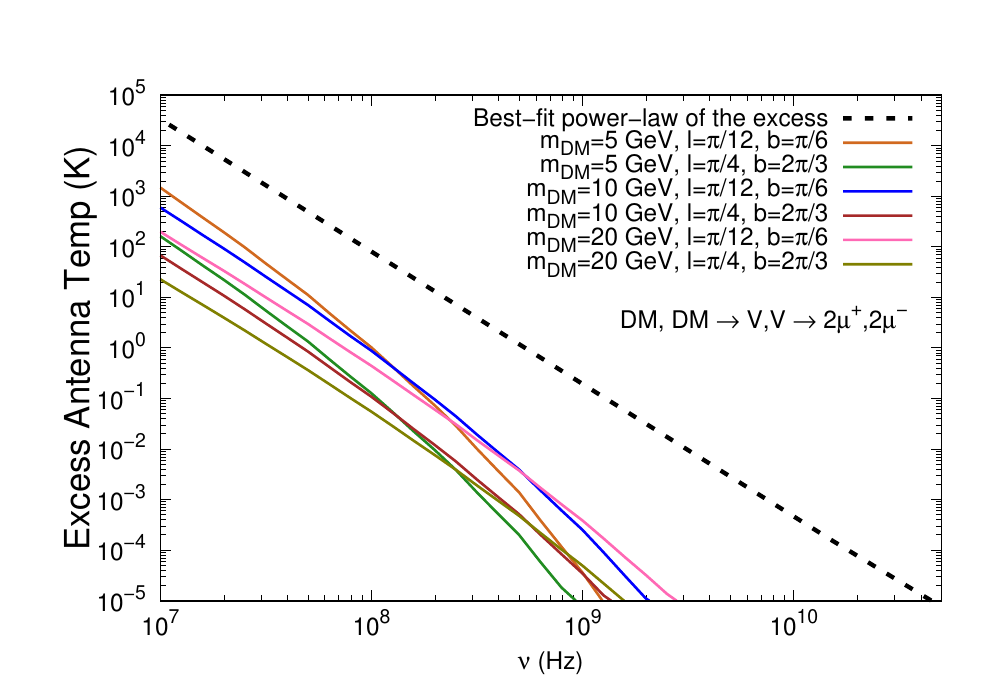}\\
\includegraphics[width=0.5\textwidth=100mm]{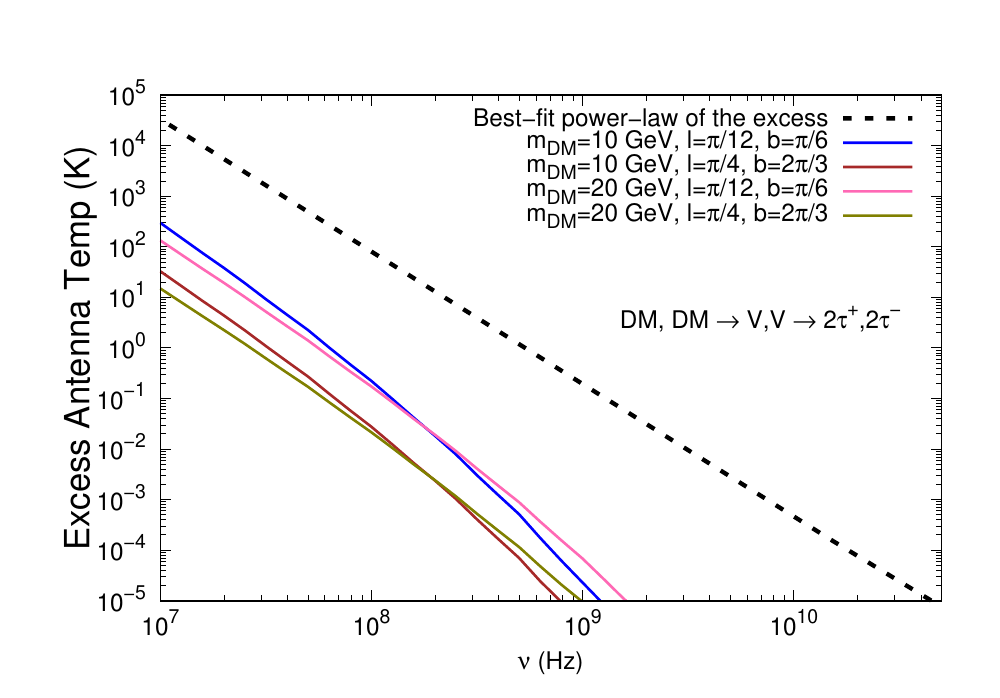}
\end{tabular}
\caption{Best−fit power−law of the excess and signatures for DM candidate annihilating through an intermediary boson. Upper left panel  shows the case $DM,DM\rightarrow V,V\rightarrow 2e^{+},2e^{-}$ while upper right panel presents $DM,DM\rightarrow V,V\rightarrow 2\mu^{+},2\mu^{-}$ for different masses (5, 10 and $20\,{\rm GeV}$). In the bottom panel we consider 10 and $20\,{\rm GeV}$ for the channel $DM,DM\rightarrow V,V\rightarrow 2\tau^{+},2\tau^{-}$ due to the kinematic constraint associated with the masses of the final particles. Plots are shown for different pairs of $l$ and $ b$ (in radians) within the ``visibility" region of ARCADE--2 experiment.  The annihilation cross section used for the simulations is  $\langle \sigma v\rangle=3\times 10^{-26}$ cm$^{3}/s$.}%
        \label{fig:e1}%
\end{figure*}

In considering the latitude and longitude curves of sensitivity of ARCADE--2 instrument, we repeated the analysis of $DM,DM\rightarrow V,V\rightarrow$ 4-lepton channels. The panels in Figure \ref{fig:e1} show our results. We can see that within the range in frequencies from $20\,{\rm MHz}$ to $\sim 5\,{\rm GHz}$ the signatures of these channels follow the power law of the radio excess, although the signal level is below that measured by ARCADE--2. In particular, the channels that provide the best results for the radio excess are $DM,DM\rightarrow V,V\rightarrow 2e^{+},2e^{-}$ and $DM,DM\rightarrow V,V\rightarrow 2\mu^{+},2\mu^{-}$.

\begin{figure*}[ht!]
\centering
\begin{tabular}{cc}
\includegraphics[width=0.5\textwidth=100mm]{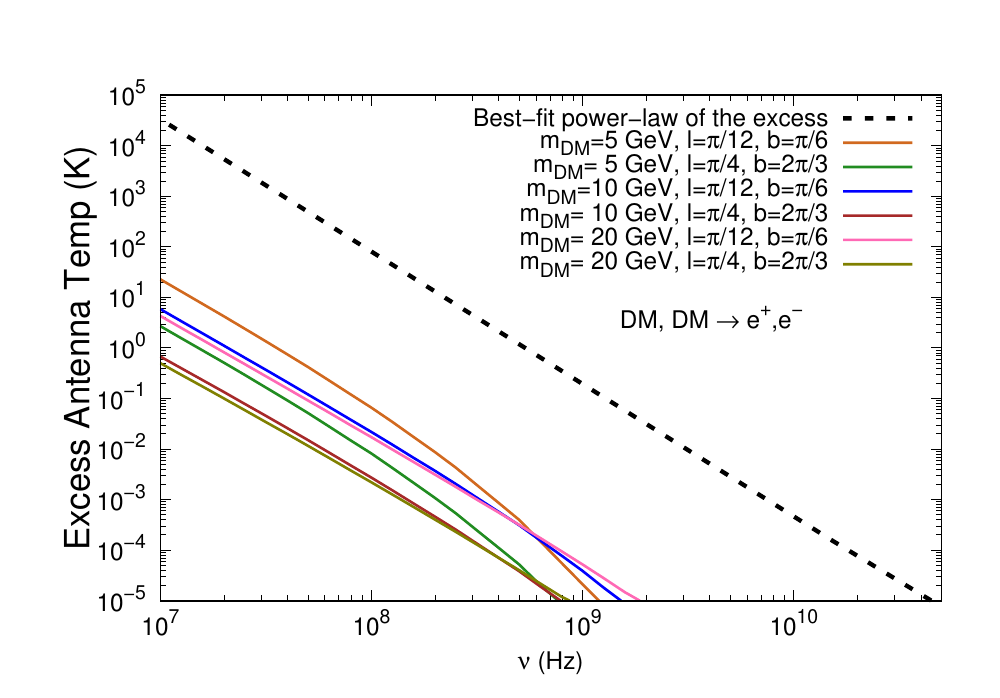}
\includegraphics[width=0.5\textwidth=100mm]{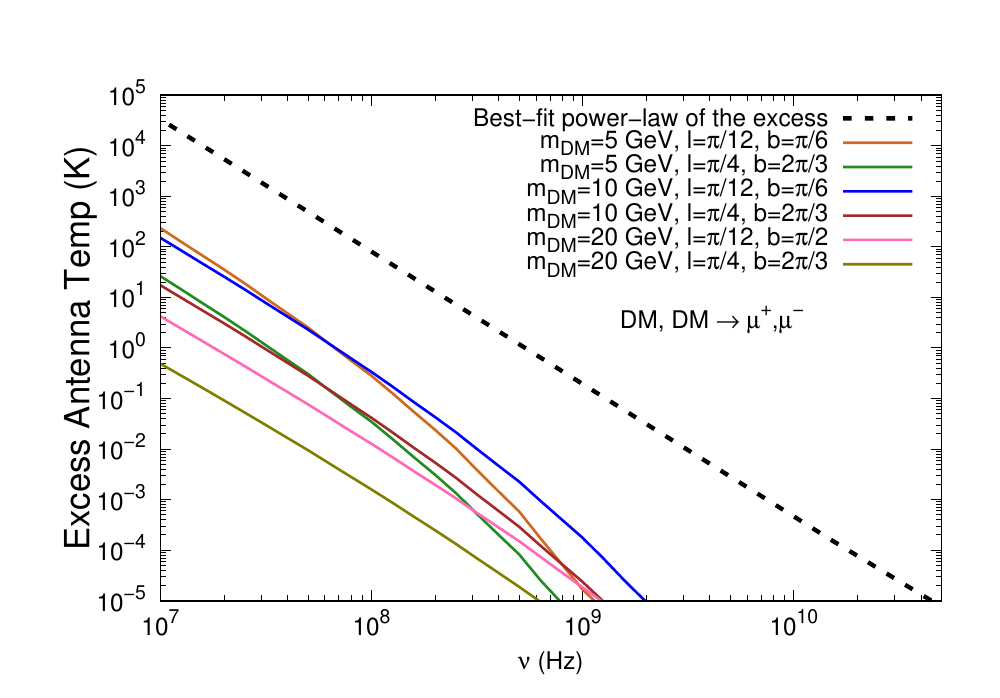}\\
\end{tabular}
\caption{Best−fit power−law of the excess compared to the signatures for DM candidate with masses 5, 10 and $20\,{\rm GeV}$ annihilating directly in $e^{+}, e^{-}$ (left panel) and $\mu^{+}, \mu^{-}$ (right panel) in a different set of latitude and longitude (in radians) within the region observed by ARCADE--2 experiment.}
    \label{fig:ex4}
\end{figure*}

We will put aside  the channel $DM,DM\rightarrow V,V\rightarrow$ four-leptons for while to focus our attention  on DM annihilation into direct leptonic channels, for  comparison purposes. The direct leptonic channels were the most analyzed in the literature in attempt to fit the ARCADE--2 excess \cite{Fornengo:2011cn}. If we consider the latitude and longitude curves of sensitivity of ARCADE--2 instrument and perform the analysis of $DM,DM\rightarrow e^{+},e^{-}$ and  $DM,DM\rightarrow \mu^{+},\mu^{-}$ channels, we will also see that the data are too far the best fit power-law excess. This can be seen in Figure \ref{fig:ex4}. For the $e^{+},e^{-}$ annihilation cross section we had considered   $\langle \sigma v\rangle=1\times 10^{-28}$ cm$^{3}\,{\rm s}^{-1}$ for $m_{DM} = 5$ and $10$ GeV and we had considered $\langle \sigma v\rangle = 9\times 10^{-28}$ cm$^{3}\,{\rm s}^{-1}$ for $m_{DM} = 20$ GeV.  For the $\mu^{+},\mu^{-}$ annihilation cross section we had considered   $\langle \sigma v\rangle=9\times 10^{-28}$ cm$^{3}\,{\rm s}^{-1}$ for $m_{DM }= 5$ and $10$ GeV and  $\langle \sigma v\rangle = 1\times 10^{-27}$ cm$^{3}\,{\rm s}^{-1}$ for $m_{DM} = 20$ GeV.

Regarding to the process $DM,DM \rightarrow V,V\rightarrow$ four leptons, we can note from the comparison of the curves in Figures \ref{fig:e1} and \ref{fig:ex4}, that the inclusion of a light intermediate vector boson has interesting implications for the radio signature resulting from the annihilation of dark matter. Firstly, it provides a stronger radio signature than direct annihilation into leptons. Secondly, the signature gives a better power-law fit to the radio excess up to frequencies $\sim 5\,{\rm GHz}$.

\section{Final Remarks}
\label{sec:Conclusion}

In this work we have studied the radio spectrum generated by DM annihilation in our galaxy. In our modeling, we have considered light DM candidates with $5 < m_{DM} < 20$ GeV annihilating into light intermediate vector bosons with the final state being a four-lepton channel.
Our choice for the mass range of the DM is associated with the channel chosen in this paper ($DM,DM\rightarrow V,V$) and with the limits established by Planck CMB observations (see, in particular \cite{pl_1} and \cite{pl_2}).

We assumed that the dark matter halo of our Galaxy follows the so-called NFW profile. We used canonical values from the literature to estimate the radio signature, through synchrotron emission, from the $DM,DM\rightarrow V,V$ channel. The radio signature obtained through the annihilation of DM near the plane of the galactic disk is shown to be in good agreement with the radio excess inferred by ARCADE--2, in combination with experiments at low frequencies. Our results show that the $DM,DM\rightarrow V,V$ signature follows the best-fit power-law of the radio excess from $20\,{\rm MHz}$ up to $\sim 5\,{\rm GHz}$, which is close to the frequency limit of the ARCADE's highest quality data ($\sim 10\,{\rm GHz}$).

In this work we have also presented, for the first time, an analysis performed with the set of latitudes and longitudes used in ARCADE--2 experiment. We also added a new analysis on the sensitivity of ARCADE--2 within the region of the sky that it observed.

Our results show that, in principle, this annihilation channel is an interesting way to explain the radio excess. However, could this signal be measured by ARCADE--2? To address this issue, we reanalyzed the ARCADE--2 data on the three frequencies at which this instrument provided the best results. Our analysis has produced new results not yet published and, in particular, allows us to show the sensitivity within the area of the sky covered by this instrument.

We verified that the annihilation of DM in the region covered by ARCADE--2 has a signature that follows the radio excess but produces a signal level below that which was measured. The channels that are most interesting are those of low mass and with annihilation through $DM,DM\rightarrow V,V \rightarrow e^{+},e^{-}$ and $DM,DM\rightarrow V,V \rightarrow \mu^{+},\mu^{-}$ channels. The inclusion of an intermediate vector boson produces a stronger radio signature at higher frequencies than the ``direct" leptonic signatures studied previously.  Although the excess signal measured by ARCADE--2 cannot be explained by this channel, the inclusion of a light intermediate vector boson is a potentially interesting channel to consider in the analysis of possible candidates for dark matter because it provides more radio power than direct annihilation in leptons.

\section*{Acknowledgments}
E. C. F. S. Fortes thanks Marco Cirelli for very useful discussion  and also acknowledges the hospitality of CERN theoretical group where part of this work was done. E. C. F. S. Fortes and O. D. Miranda thanks FAPESP for financial support under contract 2018/21532-4.


\begin{thebibliography}{99}

\bibitem{Rubin:1982}
Rubin, V.~C. and Thonnar, N.~T. and Ford, Jr., W.~K.
AJ\ {\bf 87}, 477 (1982).

\bibitem{Planck2018}
N.~Aghanim {\it et al.} [Planck Collaboration]
arXiv:1807.06209 [astro-ph.CO] .

\bibitem{Clowe_2006}
  D.~Clowe, M.~Brada{\v{c}}, A.H.~Gonzalez, M.~Markevitch, S.W.~Randall, C.~Jones and D.~Zaritsky
  The Astrophysical Journal {\bf 648}, L109 (2006)

\bibitem{Massey:2007wb}
  R.~Massey {\it et al.},
  Nature {\bf 445}, 286 (2007)
  doi:10.1038/nature05497
  [astro-ph/0701594].

\bibitem{Aprile:2019}
E.~Aprile {\it et al.} [XENON Collaboration],
Phys.\ Rev.\ Letters {\bf 122}, 141301 (2019)


\bibitem{Akerib:2016vxi}
D.~S.~Akerib {\it et al.} [LUX Collaboration],
\textrm{Results from a search for dark matter in the complete LUX exposure},
  Phys.\ Rev.\ Lett.\  {\bf 118}, no. 2, 021303 (2017).

\bibitem{Fu:2016ega}
C.~Fu {\it et al.} [PandaX-II Collaboration],
\textrm{Spin-Dependent Weakly-Interacting-Massive-Particle--Nucleon Cross Section Limits from First Data of PandaX-II Experiment},
Phys.\ Rev.\ Lett.\  {\bf 118}, no. 7, 071301 (2017)

\bibitem{Bringmann:2012ez}
  T.~Bringmann and C.~Weniger,
  Phys.\ Dark Univ.\  {\bf 1}, 194 (2012)
  doi:10.1016/j.dark.2012.10.005
  [arXiv:1208.5481 [hep-ph]].

  \bibitem{Bringmann:2012vr}
  T.~Bringmann, X.~Huang, A.~Ibarra, S.~Vogl and C.~Weniger,
  JCAP {\bf 1207}, 054 (2012)
  doi:10.1088/1475-7516/2012/07/054
  [arXiv:1203.1312 [hep-ph]].

  \bibitem{Ishiwata:2008qy}
  K.~Ishiwata, S.~Matsumoto and T.~Moroi,
  Phys.\ Rev.\ D {\bf 79}, 043527 (2009)
  doi:10.1103/PhysRevD.79.043527
  [arXiv:0811.4492 [astro-ph]].

\bibitem{Buch:2015iya}
  J.~Buch, M.~Cirelli, G.~Giesen and M.~Taoso,
  JCAP {\bf 1509}, no. 09, 037 (2015)
  doi:10.1088/1475-7516/2015/9/037
  [arXiv:1505.01049 [hep-ph]].


  \bibitem{Cirelli:2010xx}
  M.~Cirelli {\it et al.},
  JCAP {\bf 1103}, 051 (2011)
  Erratum: [JCAP {\bf 1210}, E01 (2012)]
  doi:10.1088/1475-7516/2012/10/E01, 10.1088/1475-7516/2011/03/051
  [arXiv:1012.4515 [hep-ph]].

\bibitem{Hooper:2011}
D. Hooper \& T. Linden, Phys.\ Rev.\ D {\bf 84}, 123005 (2011).

\bibitem{Ajello:2016}
M. Ajello et al. Astrophys. J. {\bf 819}, 44 (2016).

\bibitem{Ackermann:2015}
M. Ackermann et al., Phys.\ Rev.\ Letters {\bf 115}, 231301 (2015).

\bibitem{DiMauro:2019}
M. DiMauro et al., Phys.\ Rev.\ D {\bf 99}, 123027 (2019).

\bibitem{Unsold}
    A.~Unsöld, Zeits. f. Astrophysik \textbf{26}, 176 (1949).
  M.~ Ryle, Proc. Phys. Soc. London \textbf{A62}, 483 (1949).
  H.~Alfvén an N. Herlofson, Phys. Rev. \textbf{78}, 616 (1950).

\bibitem{Richard}
 Richard Wielebinski (2006) Technical Report: History of Synchrotron Radiation in Astrophysics, Synchrotron Radiation News, 19:5, 4-9, DOI: 10.1080/08940880600978663

\bibitem{Agrupados}
C.~G.~T.~Haslam, H.~Stoffel, S.~Kearsey, J.~L.~Osborne and S.~Phillips,
Nature {\bf 289}, 470 (1981).
doi:10.1038/289470a0;

R.~S.~Roger, C.~H.~Costain, T.~L.~Landecker and C.~M.~Swerdlyk,
Astron.\ Astrophys.\ Suppl.\ Ser.\  {\bf 137}, 7 (1999)
doi:10.1051/aas:1999239
[astro-ph/9902213];

\bibitem{Fixsen:2009xn}
D.~J.~Fixsen {\it et al.},
Astrophys.\ J.\  {\bf 734}, 5 (2011)
doi:10.1088/0004-637X/734/1/5


\bibitem{Fornengo:2011cn}
N.~Fornengo, R.~Lineros, M.~Regis and M.~Taoso,
Phys.\ Rev.\ Lett.\  {\bf 107}, 271302 (2011)
doi:10.1103/PhysRevLett.107.271302
[arXiv:1108.0569 [hep-ph]].

 \bibitem{Cirelli:2016mrc}
M.~Cirelli and M.~Taoso,
JCAP {\bf 1607}, no. 07, 041 (2016)
doi:10.1088/1475-7516/2016/07/041
[arXiv:1604.06267 [hep-ph]].


  \bibitem{ArkaniHamed:2008qn}
  N.~Arkani-Hamed, D.~P.~Finkbeiner, T.~R.~Slatyer and N.~Weiner,
  Phys.\ Rev.\ D {\bf 79}, 015014 (2009)
  doi:10.1103/PhysRevD.79.015014
  [arXiv:0810.0713 [hep-ph]].


  \bibitem{Wechakama:2013hra}
  M.~Wechakama,
  ``Multi-messenger constraints and pressure from dark matter annihilation into electron-positron pairs.''

  \bibitem{Kappl:2014hha}
  R.~Kappl and M.~W.~Winkler,
  JCAP {\bf 1409}, 051 (2014)
  doi:10.1088/1475-7516/2014/09/051
  [arXiv:1408.0299 [hep-ph]].

  \bibitem{Kappl:2015bqa}
  R.~Kappl, A.~Reinert and M.~W.~Winkler,
  JCAP {\bf 1510}, 034 (2015)
  doi:10.1088/1475-7516/2015/10/034
  [arXiv:1506.04145 [astro-ph.HE]].

  \bibitem{Fortes:2015qka}
  E.~C.~F.~S.~Fortes, V.~Pleitez and F.~W.~Stecker,
  Astropart.\ Phys.\  {\bf 74}, 87 (2016)
  doi:10.1016/j.astropartphys.2015.10.010
  [arXiv:1503.08220 [hep-ph]].

  \bibitem{Pospelov:2007mp}
  M.~Pospelov, A.~Ritz and M.~B.~Voloshin,
  Phys.\ Lett.\ B {\bf 662}, 53 (2008)
  doi:10.1016/j.physletb.2008.02.052
  [arXiv:0711.4866 [hep-ph]].

\bibitem{Cholis:2008vb}
  I.~Cholis, L.~Goodenough and N.~Weiner,
  Phys.\ Rev.\ D {\bf 79}, 123505 (2009)
  doi:10.1103/PhysRevD.79.123505
  [arXiv:0802.2922 [astro-ph]].

  \bibitem{ArkaniHamed:2008qp}
  N.~Arkani-Hamed and N.~Weiner,
  JHEP {\bf 0812}, 104 (2008)
  doi:10.1088/1126-6708/2008/12/104
  [arXiv:0810.0714 [hep-ph]].

  \bibitem{Elor:2015tva}
  G.~Elor, N.~L.~Rodd and T.~R.~Slatyer,
  Phys.\ Rev.\ D {\bf 91}, 103531 (2015)
  doi:10.1103/PhysRevD.91.103531
  [arXiv:1503.01773 [hep-ph]].


  \bibitem{Klypin:2014kpa}
  A.~Klypin, G.~Yepes, S.~Gottlober, F.~Prada and S.~Hess,
  Mon.\ Not.\ Roy.\ Astron.\ Soc.\  {\bf 457}, no. 4, 4340 (2016)
  doi:10.1093/mnras/stw248
  [arXiv:1411.4001 [astro-ph.CO]].

\bibitem{Floyd}
 F.~W. Stecker,
 APSS {\bf 3 }, 579  (1969)
 doi:10.1007/BF00704862

 \bibitem{Fornengo:2011iq}
  N.~Fornengo, R.~A.~Lineros, M.~Regis and M.~Taoso,
  JCAP {\bf 1201}, 005 (2012)
  doi:10.1088/1475-7516/2012/01/005
  [arXiv:1110.4337 [astro-ph.GA]].

\bibitem{Stecker:1971}
F.W.~Stecker, Cosmic Gamma Rays (book) NASA SP-249 (1971)


 \bibitem{NFW}
  J.F. Navarro, C.S. Frenk and S.D.M. White,
  Astrophys. J. 462, 563 (1996).

   \bibitem{Cirelliweb}
 {http://www.marcocirelli.net/PPPC4DMID.html}

  \bibitem{Singal:2017}
  J.~Singal {\it et al.}, arXiv:1711.09979 [astro-ph.CO]

  \bibitem{Boudaud:2016jvj}
  M.~Boudaud {\it et al.},
  Astron.\ Astrophys.\  {\bf 605}, A17 (2017)
  doi:10.1051/0004-6361/201630321
  [arXiv:1612.03924 [astro-ph.HE]].

  \bibitem{Fang}
  K.~Fang and T.~Linden,
  JCAP {\bf 1610}, no. 10, 004 (2016)
  doi:10.1088/1475-7516/2016/10/004/, 10.1088/1475-7516/2016/10/004
  [arXiv:1506.05807 [astro-ph.HE]].

  \bibitem{Barman_2017}
	B.~Barman, S.~Bhattacharya, S.~K.~Patra and J.~Chakrabortty,
	JCAP {\bf 2017}, no. 12, 021 (2017)
	doi:10.1088/1475-7516/2017/12/021
	[arXiv:1704.04945 [hep-ph]].
	
	\bibitem{pl_1}
	 M.~S.~Madhavacheril, N.~Seghal and T.~R.~Slatyer,
	 Phys. Rev. D {\bf 89}, 103508 (2014)
	
	 \bibitem{pl_2}
	 Planck Collaboration, Preprint (2015) (arXiv:1502.01589 [astro-ph.CO]) Planck 2015 Results. XIII
	
  \end{thebibliography}
\end{document}